# Infrared light scattering and emission from epsilon near-zero hyperbolic phonon-polaritons in two-dimensional crystals

Flávio Henrique Feres[1, a], Martin Bearzatto[2], Adrien Debacq[2], Gergely Nemeth[3], Rafael A. Mayer[1], Yuri B. Marçal[1, 4], Marcus V. de Paiva[1,5], Maximilian Obst[6], Felix Kaps[6], Jakob Wetzel[6], Osama Hatem[7], Wenjun Zheng[8], Ran Jing[8], Heng Wang[8], Adrian Cernescu[9], Stephan Winnerl[10], J. Michael Klopf[10], Mengkun Liu[8], Ferenc Borondics[3], Ingrid D. Barcelos[1], Raul O. Freitas[1], Ana I. F. Tresguerres-Mata[12], Yannick De Wilde[12], Lukas M. Eng[6, 13], Susanne C. Kehr[6], Michaël Lobet[2, b] & Francisco C. B. Maia[1, c]

1 Brazilian Synchrotron Light Laboratory (LNLS), Brazilian Center for Research in Energy and Materials (CNPEM), 13083-970, Campinas, São Paulo, Brazil
2 University of Namur, Department of Physics and Namur Institute of Structured Material, 61 rue de Bruxelles, 5000 Namur, Belgium
3 SMIS Beamline, SOLEIL Synchrotron, L'Orme des Merisiers, RD128, 91190 Saint Aubin, France
4 Physics Department, Gleb Wataghin Physics Institute, University of Campinas (Unicamp), Campinas, SP, Brazil
5 Department of Physics, Federal University of São Carlos 13565-905, Sao Carlos, SP, Brazil
6 Institute of Applied Physics, Technische Universität Dresden, Dresden, Germany
7 Departament of Engineering Physics, Faculty of Engineering, Tanta University, Tanta 31511, Egypt
8 Department of Physics and Astronomy, Stony Brook University; Stony Brook, New York 11794, USA
9 attocube systems GmbH, 85540, Haar-Munich, Germany.
10 Institute of Ion Beam Physics and Materials Research, Helmholtz-Zentrum Dresden-Rossendorf (HZDR), Dresden, Germany
11 Institute of Radiation Physics, Helmholtz-Zentrum Dresden-Rossendorf (HZDR), Dresden, Germany
12 Institut Langevin, ESPCI Paris, PSL University, CNRS, Paris, France
13 Würzburg – Dresden Cluster of Excellence (EXC2147) ctd. qmat – Complexity, Topology, and Dynamics in Quantum Matter, Dresden, Germany

**Corresponding authors:**

a. flavio.feres@lnls.br;
b. michael.lobet@unamur.be;
c. francisco.maia@lnls.br

# Abstract

Two-dimensional hyperbolic crystals support high-momentum ($q$) hyperbolic phonon-polariton (HPhP) modes that strongly enhance the local density of optical states (LDOS). Such hyperbolic-enhanced LDOS (HE-LDOS) is a key property that correlates high-$q$ polaritons, inherent modes of the near field electromagnetic zone, to macroscopic optical phenomena, as light emission, that rises in the far-field electromagnetic zone. The enhancement of the LDOS in hyperbolic materials has attracted considerable attention owing to its ability to boost spontaneous emission, Purcell enhancement, radiative heat transfer, and other near-field light–matter interactions. Here, we experimentally demonstrate two manifestations of the HE-LDOS in prototypical two-dimensional crystals: (i) near-field superscattering (NSSC) and (ii) near-field thermal emission (NFTE). Using scattering-type scanning near-field optical microscopy and synchrotron infrared nano spectroscopy we show that the NSSC is a light-driven effect that occurs as a well-defined, wave-like scattering into the crystal surroundings and is, primarily, located in the epsilon near-zero range. The NSSC is reasonably captured by the theoretical LDOS, based on the fluctuation-dissipation theorem, which increases in the same spectral range, ruled by augment of the density of HPhP modes. In reciprocity, the NFTE consists of the infrared radiation emission, in the far-field electromagnetic zone, due to the HE-LDOS as a function of temperature. The emission is also explained by the same LDOS theory that describes the excitation of HPhP modes by thermal stimuli using Bose-Einstein distribution. Moreover, we show that the spectral position and bandwidth of NSSC and NFTE can be tuned through crystal thickness, following the evolution of the hyperbolic polariton dispersion and the corresponding LDOS. Furthermore, polarization-resolved measurements reveal the anisotropic nature of the emission and scattering processes, directly reflecting the crystal symmetry. Our results establish a unified framework linking optically and thermally driven LDOS-mediated phenomena in epsilon near-zero hyperbolic materials, opening new opportunities for novel research in the broad scope of photonics.

**Light scattering and thermal radiation emission are, respectively, prime electromagnetic and quantum phenomena typically observed from macroscopic objects at room temperature. In subwavelength dimensions, however, their fundamentals dramatically change due to appearance of multiple channels (degenerate resonances) leading to exotic effects like superscattering and super-Planckian emission. Here we experimentally demonstrate near-field superscattering (NSSC) and the near-field thermal emission (NFTE) from two-dimensional crystals at infrared optical frequencies, mainly, around epsilon near-zero bands. Originating from optical and thermal excitation of degenerate, high-momentum hyperbolic polaritons, both phenomena are shown to be general to the class of anisotropic 2D crystals and reveal compelling properties like coherence, thickness-dependence, well-defined polarization, temperature control and directivity. Both effects are consistently explained by the fluctuation dissipation theorem and the hyperbolic-enhanced local density of states property.**

The emerging class of two-dimensional (2D) crystals presents enhanced photonic local density of optical states (LDOS) due to supporting polariton modes of high-momentum ($q$), whose values can typically surpass those of the wavevector of light in free-space $\vec{k}_0$ ($|k_0| = {}^{2\pi}/_{\lambda_0} = 2\pi\omega$, with $\lambda_0$, the free-space wavelength, and, $\omega$, the spatial frequency) by two to three orders of magnitude[1]. In polar 2D crystals, a significant class of such modes consist of hyperbolic phonon-polaritons (HPhP) that are formed by the coupling of electromagnetic fields and polar lattice vibrations[2]. Being mostly high-$q$ modes, which inherently belong to the near-field electromagnetic zone, the nanooptics of HPhPs has been characterized [3], primarily, via near-field techniques, in a broad spectral regime spanning across mid-infrared (MIR) to far-infrared (FIR) ranges [4–7]. By manipulation of the $\omega - q$ dispersion relation that is correlated with the LDOS, major achievements for controlling the fundamental physical properties of HPhPs have been reported such as: wavelength tuning[8,9], group velocity modulation[10,11], vibrational coupling in different strength regimes[12–14], and topological transitions[15]. These phenomena are discussed for well-defined hyperbolic conditions, as described below, in spectral regions wherein the electrical permittivity tensor ($\overleftrightarrow{\varepsilon}$) magnitude of the 2D media assumes appreciable values. But hyperbolic polaritons and their corresponding LDOS have been left nearly unexplored in spectral ranges when $\overleftrightarrow{\varepsilon} \to 0$, the so-called epsilon near-zero (ENZ) range albeit exquisite optical effects[16,17] are predicted in such frequencies. For example, superscattering from multilayer nanostructures[18] and the LDOS-driven radiation emission from 2D crystals[19] have been theoretically proposed upon excitation in ENZ frequencies. In related context, also super-Planckian thermal emission[20] was theoretically anticipated owing to LDOS increase in hyperbolic bands. Accordingly, recent reports have shown far-field infrared electroluminescence[21,22] from 2D heterostructures of graphene/hBN due to emission of HPhPs in the hBN that were indirectly excited by interaction with an electron gas in the graphene created by electrical currents in out-of-equilibrium regime. It is still elusive, however, the demonstration of direct effects associating the excitation of 2D crystals' LDOS in the ENZ range of hyperbolic bands – here called hyperbolic enhanced LDOS (HE-LDOS) - in the near field in association with macroscopic optical properties manifested in the far-field zone. In this work, we present novel phenomena, driven by the HE-LDOS, which establish the fundamental bridge between near-field polaritons and far-field radiation from 2D crystals: near-field superscattering (NSSC) and near-field thermal emission (NFTE) arising at ENZ frequencies[16]. Those effects are detected by both near-field specific nanoscopy tools and far-field spectroscopy in the mid-infrared (mid-IR) polaritonic bands of two prototypical 2D crystals: α-phase molybdenum trioxide (α-$MoO_3$), which is primarily discussed herein, and hexagonal boron nitride (hBN) that is examined in the supplementary materials' section 1. Importantly, we demonstrate that the NSSC and NFTE phenomena are reciprocal and intrinsically tied to the HPhPs and crystal temperature. We discuss and experimentally demonstrate key properties of those effects, which are general for the class of 2D materials, such as thickness and temperature dependence, polarization state, emission bandwidth and coherence.

### Near-field superscattering

First, we introduce the conceptual overview of the near-field superscattering (NSSC) fundamentals that are abstracted by the **Fig. 1a** cartoon. In this mechanism, illumination at a certain $\omega$ can excite HPhP modes of different orders $l$ in the medium. Ruled by the $\omega \times q$ dispersion relation ($\omega = {}^{1}/{}_{\lambda_0}$ is the spatial frequency), the allowed modes possess momentum magnitude mainly dependant on the electromagnetic boundary conditions (substrate and superstrate properties) and the material thickness as will be described here. Under external illumination, the HPhP states are populated, leveraging a large density of high-momentum polaritonic states that gives rise to the hyperbolic-enhanced LDOS (HE-LDOS). The propagating polariton fields in the medium are eventually scattered at the crystal edges[23,24] and, subsequently, converted into radiative components outside the crystal. The near-field superscattering manifestation, therefore, stems from the resultant interference between the leaky polariton radiation and illumination components that are reflected from the crystal edges. It is noteworthy commenting that the superscattering physical concepts are met in the described mechanism (Fig. 1a) since the 2D crystal is a subwavelength medium supporting degenerate HPhP modes (Fig. 3g-i) that behave as the multi-channel resonances, breaking the scattering single channel cross section[25–27](see supplementary materials' section 2). In addition, in phononic ENZ the dielectric losses are minimal[17], thus, increasing the NSSC amplitude. Here, it will be seen in the experiments that the NSSC of the α-$MoO_3$ crystal (**Fig. 1b**) is observed as high-amplitude, wave-like patterns surrounding the material. Analogous NSSC of hBN 2D crystals is shown in the supplement materials. As this effect is inherently governed by HPhPs, let us digress on their fundamentals. In general, hyperbolic modes are supported by 2D anisotropic crystals in the cases wherein the material's electrical permittivity tensor $\overleftrightarrow{\varepsilon} = (\varepsilon_{aa}, \varepsilon_{bb}, \varepsilon_{cc})$ satisfies the condition $Re[\varepsilon_{ii}].Re[\varepsilon_{jj}] < 0$, with $i \neq j$ (here, $\overleftrightarrow{\varepsilon}$ is assumed to be a diagonal matrix)[17,28] . In this notation, $a$, $b$ and $c$ represent material's crystalline axes, with $a$ and $b$ conventionally set in the basal plane and with $c$ being orthogonal to this plane (**Fig. 1c**). The hyperbolic condition allows that $q$ can have indeterminably large values for a given illumination frequency $\omega$. This is visualized from the material's isofrequency surface in momentum space that can assume a type I hyperbolic shape for $Re[\varepsilon_{cc}] < 0$ and $Re[\varepsilon_{aa}] > 0$ (or $Re[\varepsilon_{bb}] > 0$), and a type II hyperbolic shape for $Re[\varepsilon_{cc}] > 0$ and $Re[\varepsilon_{aa}] < 0$ (or $Re[\varepsilon_{bb}] < 0$)[17,29]. In the case of 2D phononic materials, such a hyperbolic condition is fulfilled within the so-called Reststrahlen bands (RB) that are defined in-between the transverse ($\omega^{TO}$) and longitudinal optical phonon ($\omega^{LO}$) resonances: $\omega^{TO} \leq \omega \leq \omega^{LO}$.

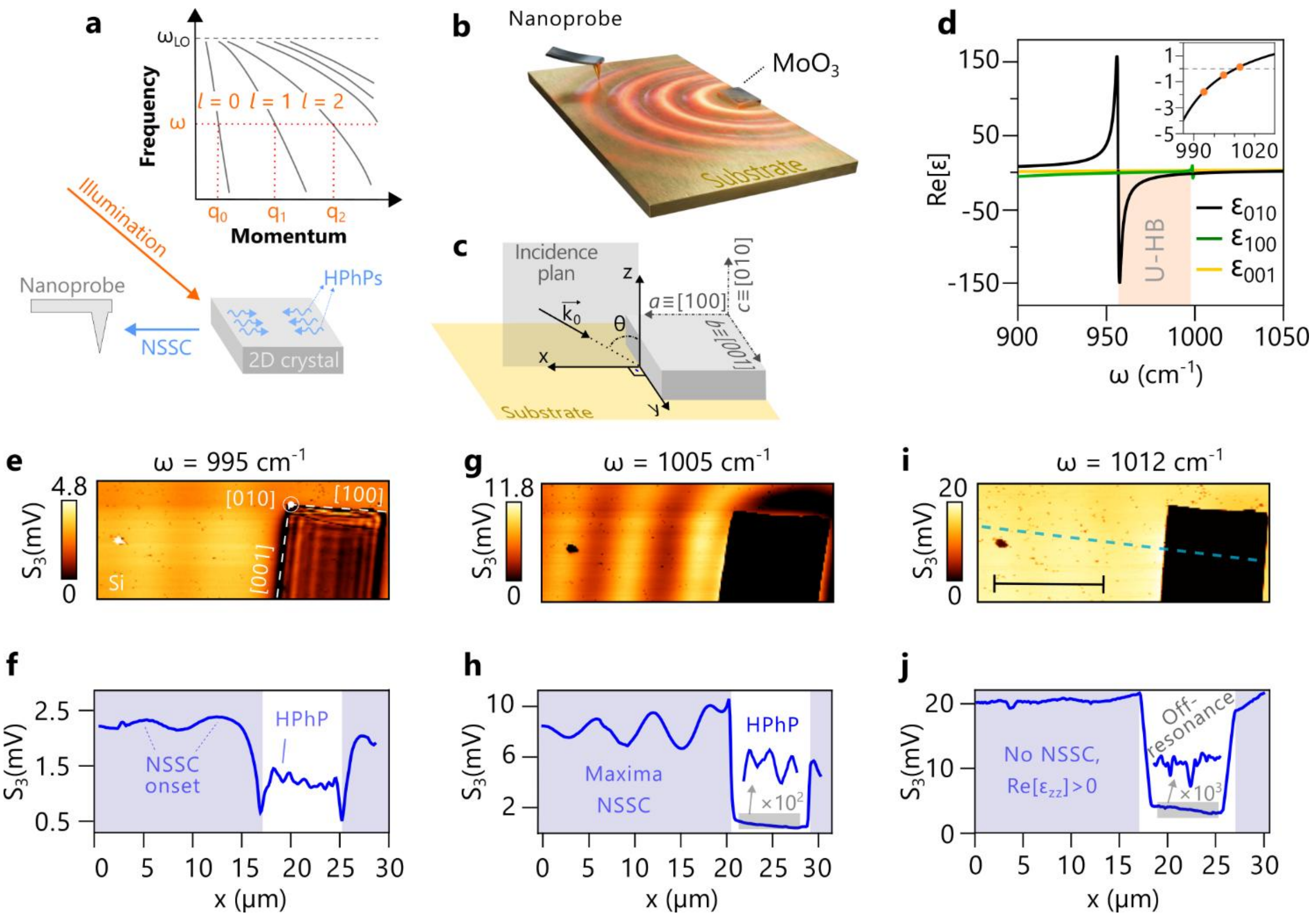

**Figure 1 - a**) Illustrative mechanism of the HE-LDOS and NSSC effect based on the frequency momentum dispersion relation. **b**) Sketch of the experimental detection of the NSSC from a α-$MoO_3$ crystal. **c**) Scheme of the relative orientation of the free-space wavevector with respect to the laboratory reference frame defined by the axes $x$, $y$ and $z$. In the experiments, the crystalline axes $a$, $b$ and $c$ are, respectively, aligned collinearly with the laboratory axes $x$, $y$ and $z$. **d**) Real part of the electrical permittivity ($Re[\varepsilon]$) of α-$MoO_3$ exhibiting the upper hyperbolic band (U-HB) with the inset depicting the frequency values associated to the measurements in **e**, **g** and **i**. s-SNOM images of a 200 nm-thick α-$MoO_3$ onto Si substrate showing (**e**) NSSC onset and HPhPs at 995 $cm^{-1}$, (**g**) NSSC maximum at 1005 $cm^{-1}$ and (**i**) an off-resonant response at 1012 $cm^{-1}$. The 10 µm scale bar in **i** is the same for **e** and **g**. Correspondingly, **f**, **h** and **j** are spatial profiles extracted from the **e**, **g** and **i**, respectively, at the direction marked by the dashed line in **i**. For the experimental images, the laser power was set around 1 mW.

As shown in **Fig. 1b**, we employ a nanoprobe of a scattering-scanning near-field optical microscope (s-SNOM) - a near-field specific instrument able to acquire narrowband nanoimages with ~ 30 nm spatial resolution[30] (see methods) – to detect the superscattering effect from a 200 nm-thick α-$MoO_3$ lying on silicon (Si) substrate. In these experiments, a quantum cascade laser (QCL) is used as illumination source with the incidence direction set to have its in-plane component orthogonal to the [001] crystalline axis (**Fig. 1c**). The QCL excitation frequencies are tuned within the mid-IR type I hyperbolic band (HB) of $\alpha$-$MoO_3$ (**Fig. 1d**), here called upper-HB. The s-SNOM image at 995 $cm^{-1}$ (**Fig. 1e**) reveals the onset of the NSSC in the crystal surroundings and the characteristic HPhPs belonging to the $\alpha$-$MoO_3$ U-HB in the medium. The spatial limits of these effects can be visualized from the spatial profile shown in **Fig. 1f**. The HPhPs form the expected standing wave feature inside the material, where one finds maxima periods of $\lambda_p = 2\pi/q \sim 1.6$ µm. The NSSC, nevertheless, is found outside of the $\alpha$-$MoO_3$ crystal with maxima period of, $\lambda_{sc} = 8.4$ µm, which is considerably larger ($\lambda_{sc}/\lambda_p \sim 5$) than those of the HPhPs inside the crystal. For the 1005 $cm^{-1}$ illumination frequency (**Fig. 1g**), the NSSC reaches its maxima of intensity, whilst the signal on the crystal is minimal since this frequency approaches the ENZ range of the material where the s-SNOM signal amplitude is an asymptote near zero. This occurs because a sufficiently extended ENZ medium does not support electromagnetic waves due to its high impedance[17,19], like a metal near the plasma frequency. Note that we are considering the ENZ range of $\overleftrightarrow{\varepsilon}$ in [010] crystalline axis (**Fig. 1d**). The 1005 $cm^{-1}$ profile (**Fig. 1h)** indicates the strong presence of the NSSC outside the crystal, whilst a HPhP standing wave ($\lambda_p \sim 3.7$ μm) can still be visualized in the medium upon multiplying the profile curve values by a factor of $10^2$. While ENZ-driven confinement of phonon-polaritons has been recently described from spectroscopy view[31], the HPhP wave shown in **Fig. 1h** is the first experimental demonstration of genuine ENZ HPhP propagating modes. Such polariton experimental $q$'s approximately match the theoretical dispersion relation for HPhP's (see supplementary materials' section 3**)**. Neither the NSSC nor HPhP appear at 1012 $cm^{-1}$ (**Fig. 1i**), even by using a $10^3$-multiplication factor to amplify the signal from the medium (**Fig. 1j**). We label this behaviour as off resonance. Importantly, the NSSC manifests itself in the end but within the U-HB of $\alpha$-$MoO_3$ (**Fig. 1d**) where $Re[\varepsilon_{[010]}]$ has small, negative values (see **Fig. 1d** inset), close to zero. The off-resonance image, however, happens when $Re[\varepsilon_{[010]}]$ becomes positive, albeit with small values still. As general note, in the U-HB, $\varepsilon_{[010]}$ is resonant with [010] lattice vibrations leading to negative values of its real part in between $\omega_{010}^{TO} = 956.7$ $cm^{-1}$ and $\omega_{010}^{LO} = 1006.9$ $cm^{-1}$, whereas $Re[\varepsilon_{[001]}]$ and $Re[\varepsilon_{[100]}]$ have both only positive. Notwithstanding the fact that $\alpha$-$MoO_3$ possesses a rich polaritonic activity throughout the far- to the mid-infrared range[32] (see supplementary materials' section 4), we primarily discuss the NSSC of the upper-HB ($Re\left[\varepsilon_{[010]}\right] < 0$ a $Re\left[\varepsilon_{[100]}\right]; Re\left[\varepsilon_{[001]}\right] > 0$) (**Fig. 1d**) for the near-field measurements, since it is largely dominating over the analogue effect observed in the other bands (supplementary materials' section 4). As will be elucidated farther in **Fig. 4**, this is a direct consequence of the specific polarization states of the NSSC for each HB. In the U-HB, the NSSC is polarized parallel to the nanoprobe main axis, hence, its scattering is favourably selected in the near-field measurements. In opposite, the NSSC of the type II bands has polarization orthogonal to the nanoprobe main axis, thus, being less efficiently detected.

To spectrally characterize the NSSC, we used synchrotron infrared nanospectroscopy (SINS) that enabled acquiring the spatio-spectral linescan across the edge of a 128 nm-thick $\alpha$-$MoO_3$ crystal lying on gold (Au) substrate (**Fig. 2a)**. In analogy to s-SNOM, SINS can produce hyperspectral images with ~ 30 nm spatial resolution (see methods) since the excitation stems from a broadband, highly brilliant infrared beam from a synchrotron source[33,34]. In the measurement shown in **Fig. 2a**, the SINS nanoprobe scans a 30 $\mu$m-long total distance: ~ 1 μm on the $\alpha$-$MoO_3$ and 29 μm on Au surface, away from the $\alpha$-$MoO_3$ crystal where the superscattering occurs. On the crystal, there is a high amplitude signal owing to the polaritonic activity in the U-HB. From the crystal edge, in agreement with the commented s-SNOM images of **Fig. 1**, a strong wavy signal is observed, around $\omega_{010}^{LO}$ which, as discussed above, corresponds to the ENZ frequency for the out-of-plane component. The wave pattern extends over the whole scanned distance on the Au surface. Moreover, the linescan allows determining the NSSC spectral linewidth ranging from ~990 to 1010 $cm^{-1}$, with a 5 $cm^{-1}$ spectral resolution, for the measured 128 nm-thick crystal. Yet, as described below, we find that the linewidth is dependent on the crystal thickness. A detailed examination of the linewidth, for different crystal thicknesses, is provided in the supplementary materials' section 5. As discussed above, the near-field superscattering, seen in s-SNOM and SINS measurements, originates from the same interference mechanism that is sketched in **Fig. 1a**. To fully reproduce the experimental linescan (**Fig. 2a**), we call on the simulated spatial structure of the modulus of the $z$-component resultant electric field ($E_z$) as shown in **Fig. 2b**. The full-wave simulations compute, in analogy with the experiments, a monochromatic light swept over a specified range of frequencies with the same relative orientations between the incidence wavevector and the sample axes (see methods). Accordingly, the $|E_z|$ simulations reproduce the oscillatory beam that propagates from the $\alpha$-$MoO_3$ crystal edge to the Au surface. This oscillatory emission matches the spatial period as that detected in the experimental linescan as shown by the good matching between experimental and simulated profiles for $\omega = 1002$ cm$^{-1}$ (inset in **Fig. 2b**). It is noteworthy that the NSSC can be considered, by origin, as a spatially coherent effect that is inherited from the HPhPs. The well-defined field structure seen from the nanoscopy results confirms such coherence.

In the **Fig. 2c**, we study the spectral behaviour of this near-field superscattering phenomenon, around the ENZ frequency along z direction. It can be understood by inspecting the local electromagnetic energy density of the system $U(z, \omega)$, within the framework of the fluctuation-dissipation theorem (FDT) as described by Joulain *et. al* [35,36]:

$$U(r, \omega) = \Theta(\omega, T).\,\mathrm{P}(r, \omega), \qquad (1)$$

that the is product of the mean thermal energy per photon $\theta(\omega, T)$, stemming from Bose-Einstein distribution (2), and the local density of states energy $\mathrm{P}(r, \omega)$ (3):

$$\theta(\omega, T) = {hc\omega}\Big/{e^{\frac{hc\omega}{k_B T}} - 1}, \qquad (2),$$

and

$$\mathrm{P}(z, \omega) = \frac{2\omega^2}{c}\left[\int_0^{k_0} \frac{k_p}{k_0}\frac{dk_p}{|k_{ez}|}\frac{(1 - |r_s|^2) + \left(1 - |r_p|^2\right)}{2} + \int_{k_0}^{\infty} \frac{4{k_p}^3}{{k_0}^3}\frac{dk_p}{|k_{ez}|}\frac{Im[r_s] + Im[r_p]}{2} e^{-2Im[k_{ez}]z}\right] \qquad (3).$$

In eq. 2 $h$ is the Planck constant, $\omega$ is the spatial frequency as defined above, $k_B$ is the Boltzmann constant, and $T$ is the temperature. In eq. 3 $r_s$ and $r_p$ are the Fresnel reflection coefficients in $s$- and $p$-polarizations, $k_p$ and $k_{ez}$ are the in-plane and the out-of-plane (extraordinary) wavevector components, and $z$ is the distance from the material interface, respectively. A close inspection of the spectral energy density distinguishes the propagating - first integral in (3) - and evanescent – second integral in (3) - contributions. In the evanescent regime, dissipation is dominated by the polaritonic response of the system, as it is proportional to $\mathrm{Im}[r_s]$ and $\mathrm{Im}[r_p]$. Consequently, the LDOS enhancement near the interface is driven by the spatial extension of evanescent fields associated with the HPhPs discussed above. It is observed **in Fig. 2c** that the LDOS increases up to 5 orders of magnitude within the same spectral range of the near-field superscattering effect. Furthermore, analytical model, simulations and experiments consistently show that NSSC occurs within the spectral range where the HE-LDOS is predicted. The LDOS defines the number of available electromagnetic modes, while their occupation is governed by the Bose–Einstein distribution under thermal excitation. In optical measurements such as s-SNOM and SINS, these states are populated by the external illumination. Therefore, spectral behaviour of NSSC can be understood as a light-driven manifestation of the HE-LDOS, representative of the HPhP modes that are accessible under the specified synchrotron and laser excitations in the near-field experiments.

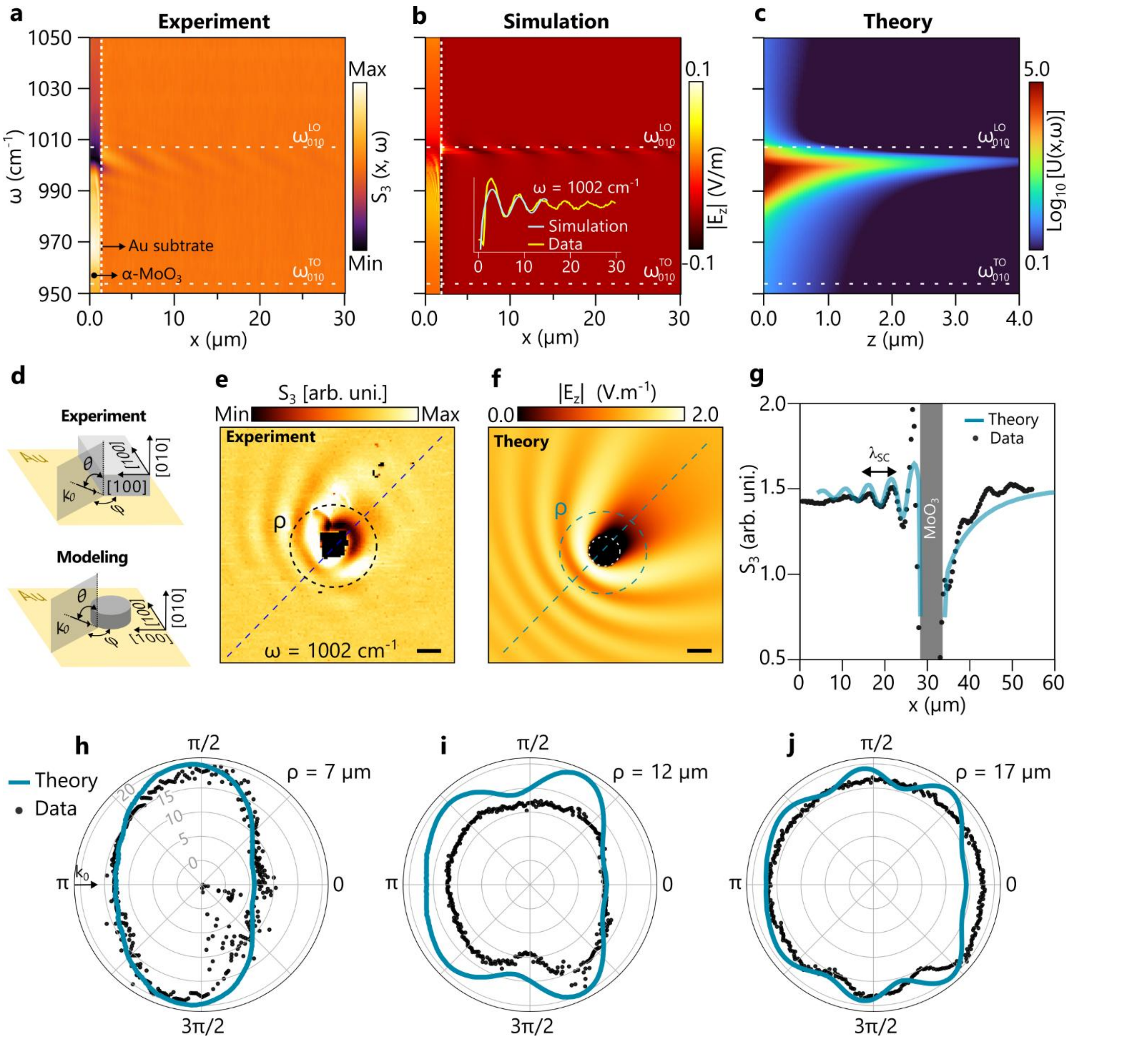

Figure 2 - **a)** 30 $\mu$m-long spatio-spectral linescan by SINS, given from the normalized third harmonic amplitude $S_3(\omega, x)$. This measurement was performed from the edge of a 128 nm-thick α-$MoO_3$, lying on Au substrate. **b)** $|E_z|$ simulation in the frequency domain. **c**) Spectral energy density/LDOS amplitude calculated from the interface of α-$MoO_3$ and vacuum for a 128 nm thick sample. **d**) Illumination incidence schemes used in the experiments (**e**) and theoretical modelling (**f**) both inspected at ω = 1002 cm$^{-1}$. Scale bar represents 5 μm. The crystal is 5 x 5 $\mu m^2$. **g**) Line profiles extracted along the dashed line in **e** and **f**, with the grey region indicating the crystal position. **h–j**, Angular distribution of the scattered intensity at radial distances ρ = 7, 12, and 17 μm. Experiment (black dots) and theory (blue lines).

Because the lateral dimensions of the crystal are comparable to the excitation wavelength, the crystal can efficiently scatter the incident radiation. Hence the combination of HE-LDOS theoretical framework with Mie scattering theory[37–39] provides a natural description of the interference patterns observed around the crystal in narrowband images (**Fig. 2e–j**). While the HE-LDOS describes the spectral behaviour, the Mie scattering provides a description for the spatial wave patterns around the crystal. In the Mie scattering theoretical analysis, we model the crystal with cylindrical symmetry to simplify the mathematical approach (see supplementary materials), where the diameter is 5 μm (**Fig. 2d**). The observed NSSC fringes therefore originate from the interference between the incident field, $E_0$, and the field scattered by the crystal, $E_M$, giving rise to the anisotropic wave-like patterns surrounding the crystal. Such simulation (**Fig. 2f**) captures the overall behaviour of the corresponding experimental data (**Fig. 2e**) within those hypotheses. Moreover, we find correct match between experimental and calculated (**Fig. 2g**) line profiles, thus further confirming the convergence between theory and experiment. We also observe that the NSSC shows high in-plane anisotropy around the crystal. On the illumination side (anti parallel to $k_0$), the NSSC pattern produces smaller spatial period of $\lambda_{sc}$. While behind the crystal a dark region (parallel to $k_0$) can be seen, and a larger $\lambda_{sc}$ is observed. Small discrepancies between theory and experiments, can be attributed to the hypothesis used in this theoretical model, which assumes a cylindrical geometry, whereas the actual crystal has a square shape and therefore a different scattering cross-section. This examination is complemented by the angular distribution of the NSSC field, for different radii $\rho$, as depicted in **Fig. 2 h-j.** where the NSSC confirms directivity, expected for Mie scattering, by varying the azimuthal angle φ (see supplementary materials' section 2). Interestingly, light scattered by ENZ media has been largely investigated in the far-field regime for macroscopical structures, wherein the ENZ media can shape the upcoming wavefront[16,39,40]. Nevertheless, the NSSC of 2D materials, as studied here, consists of the first observation of ENZ superscattering from hyperbolic 2D materials, in optical frequencies and in the near field regime.

Furthermore, the HE-LDOS has an important dependence on the crystal thickness *d*, which modulates the NSSC spectral behaviour, which we confirm here via experiment and theory. These observations are shown by SINS linescans in the vicinity of crystals with *d* = 140, 270, and 808 nm (**Fig. 3a-c**). In a qualitative view, we note that the NSSC reveals a redshift of maximum (blue arrow) of the central peak and the raise of lower intensity sidebands (red arrows) as *d* increases. To take thickness role into account, we modified equation (3), introducing thickness dependent and anisotropic Fresnel coefficients (see supplementary materials' section 9). Hence, the calculated HE-LDOS, in *z* direction, for finite thickness and anisotropic material (**Fig 3d-f**) reveals comparable spectral changes as a function of the thicknesses. Considering the modelling in eq. 1, the thickness dependence is introduced via $r_p$ (see supplementary materials' section 9). To understand this behaviour, we recall the strong correlation between the HE-LDOS and the $\omega \times q$ dispersion relation of the HPhPs. The dispersions are shown from the imaginary part of the global $r_p$, which is calculated by solving Maxwell's equations for the layered system: air/ $\alpha$-$MoO_3$/Au. In **Fig. 3g-i** we plot the calculated dispersion for the upper-HB, with respect to the polaritonic activity in the [010] - [001] plan, and for crystals of the same thicknesses. Thinner crystals tend to support flatter HPhP dispersion [17,41,42], concentrated in a narrower band and with mode density growing towards the ENZ range, while the HPhP dispersion of thicker ones has higher population of modes filling larger range of the hyperbolic band. This is seen from the dispersion of the 140 nm-thick crystal (**Fig. 3g**), the branch of the lowest order mode ($l = 0$) covers a 980-1007 $cm^{-1}$ spectral window within the momentum range from 0 to $10^5$ $cm^{-1}$, whilst the high order modes remain confined near $\omega_{010}^{LO}$. As a result, it occurs the narrowing and the spectral shift of the HE-LDOS around the ENZ frequency due to the high density of HPhP modes at this frequency [**Fig. 3d**]. This indicates that an optical flat-band dispersion leads to narrowing and enhancement of the LDOS[43,44]. As $d$ increases, however, more modes are allowed in the medium as exhibited by the dispersion of the 808 nm-thick crystal (**Fig. 3i**), wherein branches of different orders tend to cover larger spectral ranges in the U-HB (**Fig. 3h** and **i**). The contribution of such higher order modes to the HE-LDOS leads to the appearance of the broadened maximum around 990 cm$^{-1}$ for the 808 nm-thick crystal (**Fig. 3f**). Theoretical predictions indicate that the LDOS reaches an asymptotic regime for crystal thicknesses above 1000 nm (see supplementary materials' section 5). In this regime, the LDOS approaches the bulk response and spans the entire hyperbolic band[20]. Such theoretical spectral broadening is also observed in the NSSC experiments (**Fig. 3c**). The experimentally observed sidebands (red arrows in Fig. 3a–c), however, can be attributed to edge-selected scattering channels, created by the supported momentum distribution and the geometry of the medium, akin to a cavity effect[45]. Thus, scattering process is mediated by the edge-scattering efficiency that privileges the activity of certain polariton modes, consequently, producing discrete sidebands instead of a continuous spectral distribution. Therefore, this analysis indicates that the NSSC can be tuned over the U-HB as function of the thickness and shifts asymptotically to $\omega_{010}^{LO}$, i.e. $\omega_{\mathrm{ENZ}}$, for thinner crystals (see supplementary materials' section 5). It is noteworthy commenting that the HE-LDOS-driven superscattering can be considered a multifrequency superscattering, as theoretically predicted [46], since it is also observed in the other type-I and type-II hyperbolic bands of α-$MoO_3$, spanning the 400-1100 cm$^{-1}$ broadband range. The NSSC of those bands are shown from the linescan across the edge of the 808 nm-thick crystal (supplementary materials' section 5). The type II NSSC, from the thinner crystals, is less prominent yet.

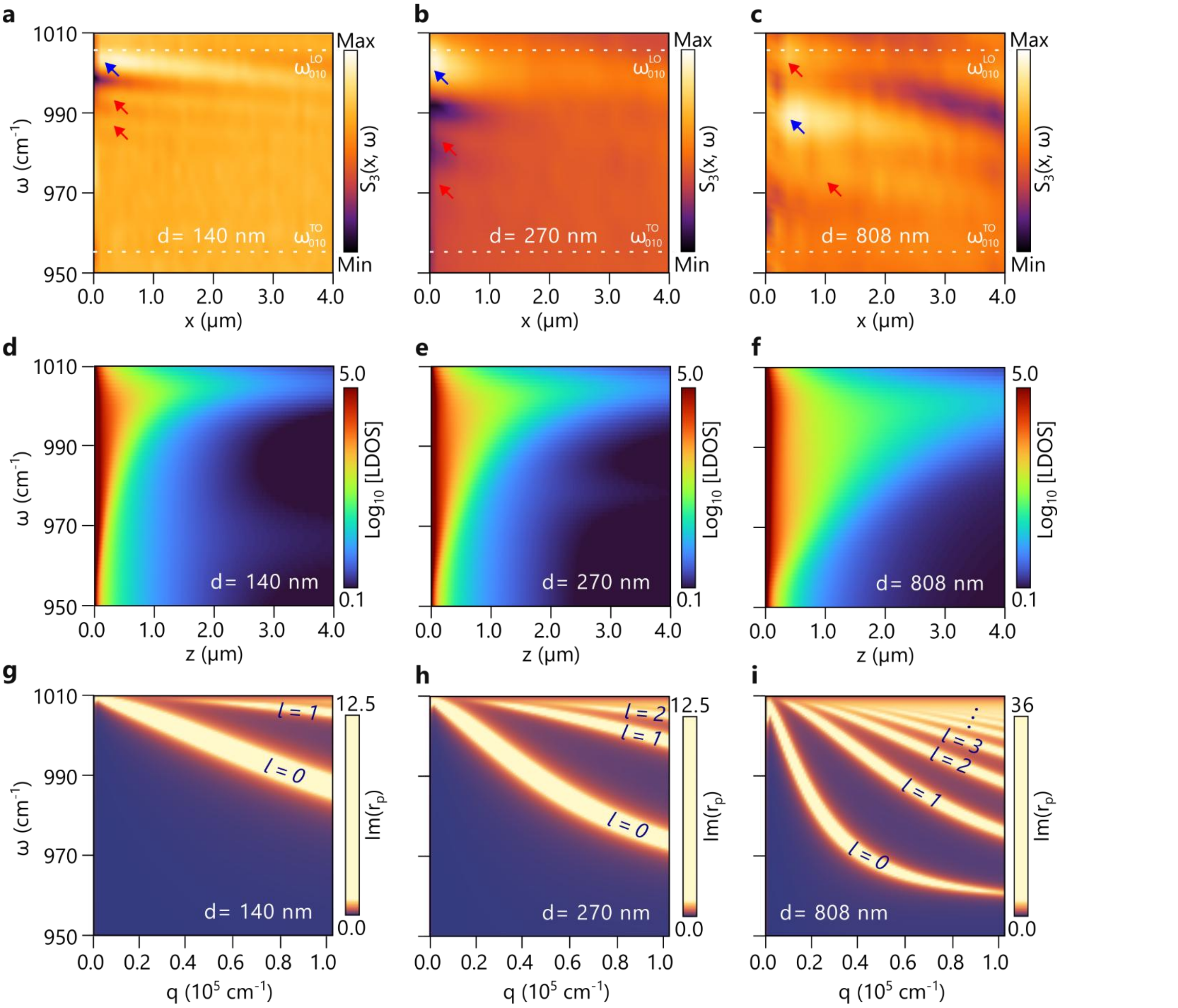

**Figure 3 – Thickness-tunable LDOS in hyperbolic crystals. a - c**) Measurements of SINS linescans shown by the normalized $S_3(\omega, x)$. (**d – f**) spectral LDOS calculated from the modified eq. (3) as a function of the distance from the sample surface $z$, and (**g-i**) calculated $\omega \times q$ dispersion relations for α-$MoO_3$ crystals with thicknesses d = 140 nm, 270 nm and 808 nm. The experiments (a-c) used a spectral resolution of 3 $cm^{-1}$. The dispersion relations are shown from the imaginary part of the $p$-polarized reflectivity coefficient $(r_p)$ and are plotted in false colour scale.

## Near-field thermal emission

Until this point, we have discussed the NSSC effect as a result of the HE-LDOS excitation by photons, considering the polariton modes activated by light excitation in the SINS and s-SNOM experiments as discussed in the **Fig.1** and **2**. However, HE-LDOS is also dependent on the temperature $T$, which appears in the Bose-Einstein distribution term $\theta(\omega,T)$ of eq. 1 and 2. Accordingly, thermal stimuli, which allow one to excite all available electromagnetic modes, can be used as parameters to modulate HE-LDOS amplitude that is predicted to increase with $T$ as shown **Fig. 4a**. For $T =$ 300, 350, 420 and 450 K, the calculations monitor variations in the HE-LDOS spectra at $x = 50\,nm$ from the [001] surface of a 150 nm-thick $\alpha$-$MoO_3$ crystal in the U-HB. We name this effect as near-field thermal emission (NFTE) since it emerges from the same framework of the NSSC as will be elucidated below. Strikingly, **in Fig. 4b** we experimentally verify these predictions by infrared spatial modulation spectroscopy (IR-SMS)[45] measurements of 150 nm-thick crystal for the different values of $T$. We should stress that IR-SMS experiments have run in complete absence of light excitation by external sources, thus, collecting only the thermal emission in the far-field electromagnetic zone by specific lock-in based detection that suppresses background contribution (see methods for details of IR-SMS). In analogy to the NSSC, the measurements with 2 $cm^{-1}$ spectral resolution, shown in **Fig. 4c**, unveil that the thermal emission is dependent on the crystal thickness: for fixed T = 450 K, the NFTE of the 500 nm-thick crystal is overall appreciably higher than that of the 150 nm-thick one. In **Fig. 4c**, we also plot also the curves of $Re[\varepsilon_{001}]$ and $Re[\varepsilon_{010}]$ to further confirm that the emission occurs within spectral ranges where they assume negative values (marked blue-shadowed region). This indicates that the thermal radiation emission occurs in the hyperbolic bands with the high-$q$ HPhP modes playing fundamental role in the radiation mechanism. Furthermore, **Fig. 4c** spectra show that the band ~ 1000 $cm^{-1}$ presents higher intensity that that ~ 800 $cm^{-1}$ for both crystal thicknesses. Considering the differences in spectral broadening due to the spectral resolution used in the experiments, theory (**Fig. 4f**) and experiment (**Fig. 4b** and **4c**) exhibit correspondence in terms of the central frequency $\Omega$ and linewidth $\Upsilon$ that are determined from a gaussian model (see supplement). From this modelling, we obtained a narrow spectral linewidth for the 2 $cm^{-1}$ spectral resolution experimental NFTE that leads to an enhanced quality factor $Q = 201$ (see supplementary materials' section 8). This indicates good correspondence with previous works on near-field thermal emission, where *quasi*-monochromatic and coherent light emission was observed in silicon carbide metasurfaces[47]. This emission linewidth is comparable to that of 2D transition metal dichalcogenides in the visible spectrum, at cryogenic temperatures[48] and electronic pumped phonon polariton lasers[49]. Note that the measurements in **Fig. 4b** and **4c** do not have polarization control that is performed in **Fig. 4e**. In the near-field thermal emission mechanism (**Fig. 4d**), the thermal excitation causes electromagnetic fluctuations that are induced due to thermal phonons, as described by the FDT[35,36,50]. These thermal fluctuations act as dipole sources that excite HPhP modes within the crystal, populating the HE-LDOS according to Bose-Einstein statistics. The high-$q$ HPhPs are (re-) scattered at the crystal edges and converted into radiation fields detectable in the far field, in close analogy with the NSSC process, therefore.

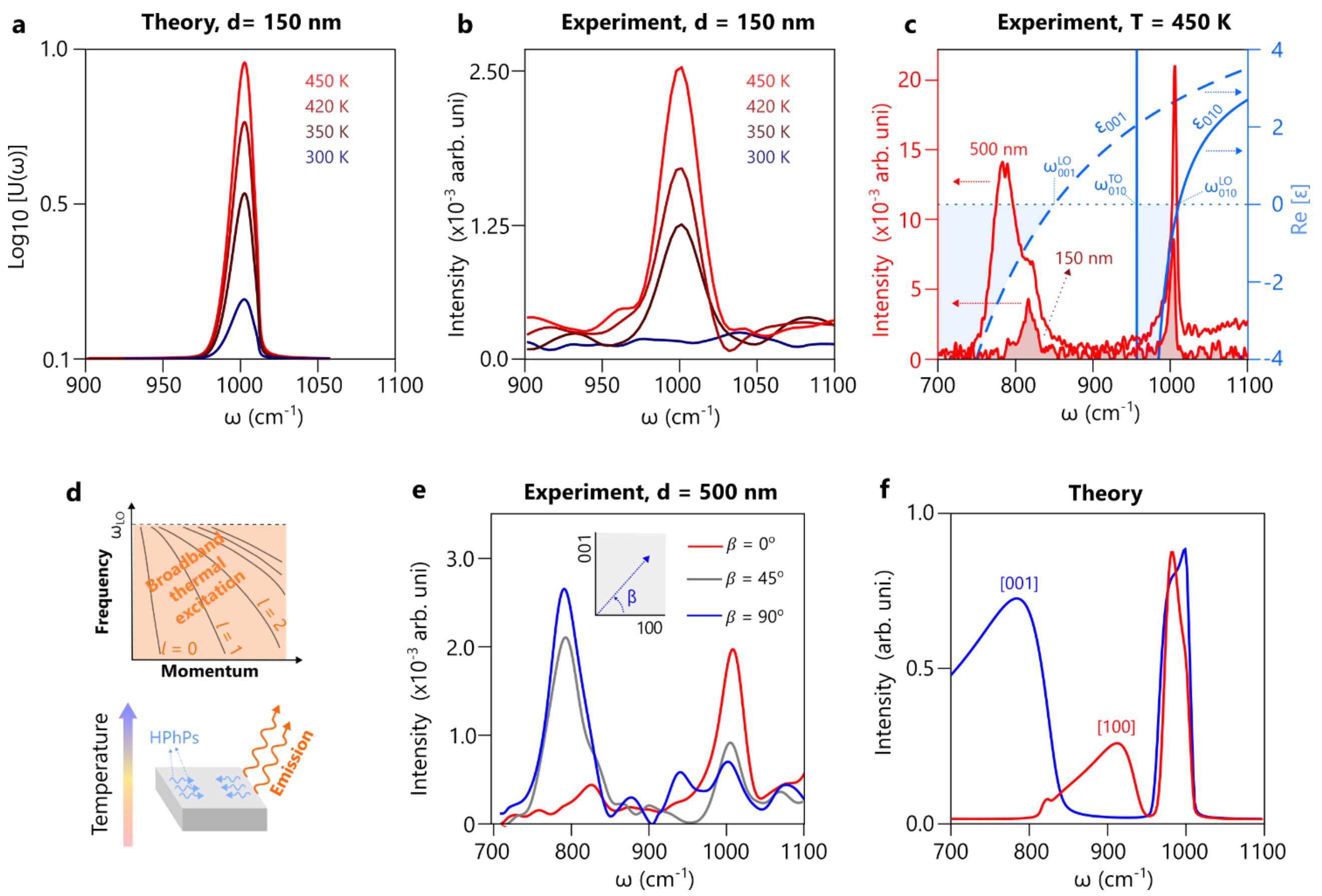

**Figure 4 –Controlled emission from NFTE of hyperbolic crystals. (a)** HE-LDOS spectra calculated of a 150 nm-thick $MoO_3$ crystal at $\boldsymbol{T}$ = 300, 350, 420, and 450 K the crystal interface. **b**) For the same temperatures, IR-SMS spectra of a 150 nm thick crystal measured with spectral resolution of 20 $cm^{-1}$. (**c**) With respect to coordinate axis on the left, IR-SMS spectra of two crystals of different thicknesses (150 and 500 nm) measured with 2 $cm^{-1}$ spectral resolution at T = 450 K. With respect to the coordinate axis on the right, **Re**[**ε**] for the [100] and [010] crystalline directions with the respective spectral locations of $\boldsymbol{\omega^{TO}}$ and $\boldsymbol{\omega^{LO}}$. **d**) Schematics of the HPhP emission mechanism due to thermal excitation. **(e)** Polarized IR-SMS spectra performed from a 500 nm-thick crystal at T = 450 K. The inset represents a schematic of the experiment where the polarization angle **β** is measured from the [100] crystalline axis and the analyser main axis (blue arrow). **(f)** HE-LDOS calculated for a 500 nm-thick crystal along the directions [001] (blue) and [100] red.

We also experimentally examine the NFTE polarization (**Fig. 4e**) using a polarization analyser that forms a relative angle $\beta$ between the analyser main axis and [100] crystalline axis of the measured crystal (**Fig. 4e** inset). The polarization-dependent measurements were performed on 500 nm-thick $\alpha$-$MoO_3$ at $T$ = 450 K (**Fig. 4e**). For $\beta$ = 0°, the NFTE spectrum shows a main peak ~ 1000 $cm^{-1}$ and a secondary band ~ 820 $cm^{-1}$ of lower intensity. Interestingly, a dramatic change is seen for $\beta$ = 90° revealing a NFTE spectrum with a more pronounced band centred ~ 780 $cm^{-1}$ accompanied by a reminiscent band ~ 1000 $cm^{-1}$. For comparison, $\beta$ = 45° spectrum (grey curve) is a convolution of the spectra from those orthogonal components. In good agreement with experimental observations, we remark that the experimentally observed polarization of the NFTE spectrum is correlated to the modified thickness-dependant and anisotropic LDOS predictions for the [100] and [001] crystal axes as shown in **Fig. 4f**. The LDOS calculated at the [001] axis presents spectral shape and resonant bands matching the measurement for $\beta$ = 90°, while the theoretical spectrum for the [100] finds comparably good analogy with the measurement for $\beta$ = 0°. We assign the measured NFTE band ~ 780 $cm^{-1}$ to the emission of type II HPhPs in the 544.6 – 850 $cm^{-1}$ that is formed due to phonon resonances polarized in the [001] crystalline axis. For this reason, such band is maximal for $\beta$ = 90° corresponding to the analyser main axis parallel to the [001] crystal axis. The 1000 $cm^{-1}$ NFTE band is nearly twice reduce from $\beta$ = 0° to 90°, since $\beta$ = 0° selects emission from both the [010]- and the [100]-polarized phonon bands, respectively, at $\omega$ = 956.7 – 1006.9 $cm^{-1}$ and $\omega$ = 998.8 – 999.2 $cm^{-1}$ (see supplement material). Otherwise, $\beta$ = 90° blocks the [100] contribution while selecting only the [010]-polarized NFTE. It is noteworthy mentioning that chiral thermal emission was recently reported from twisted heterostructures of $\alpha$-$MoO_3$ 2D crystals as an effect independent of system's hyperbolic polaritons[51]. Our study, however, demonstrates that thermal emission from single 2D crystals of $\alpha$-$MoO_3$ is directly dictated by the material's hyperbolic response, i. e. the HE-LDOS, with key importance for the high-$q$ HPhPs. Furthermore, we show that the NFTE possesses thickness dependence, linear polarized states over different phononic bands, and high $Q$ factor.

## Conclusion

In conclusion, we present here the light-driven near-field superscattering (NSSC) and the thermally activated near-field emission (NFTE) as general optical phenomena of hyperbolic two-dimensional (2D) crystals. These effects originate from HE-LDOS around the ENZ frequency, populated by electromagnetic states in the form of high-momenta hyperbolic phonon polaritons (HPhP). Our observations are experimentally derived from the mid-infrared optical response of α-$MoO_3$ and hBN 2D crystals that, as standard platforms for the study of HPhPs, can be considered representative for the class of 2D materials. While having been reported only for microwave frequencies[26], here we experimentally demonstrate the superscattering at optical frequencies, in the near-field electromagnetic zone by use of nanoscopy tools. Our observations substantially expand the reach of the superscattering[27] as the effect can now be further scrutinized within the vast class of hyperbolic materials with optical activity ranging from visible to the THz frequencies. As prime identification, the superscattering reveals a wavy-like structured optical field in the medium surroundings and its fullness occurs in the epsilon-near-zero range where the HPhP degenerate modes' density massively increases, as predicted[25]. The thickness-dependent LDOS theory, based on the fluctuation dissipation theorem, provides fundamental concepts to describe the NSSC as it computes the contribution of all degenerate hyperbolic polariton modes to explain the near-field enhancement. Importantly, this theory introduces the temperature as a variable for the superscattering, thus, unravelling an effect that can be explored in future research. Moreover, the theory also links the near-field superscattering to thermal emission in the far field that is, consistently, demonstrated as coherent infrared radiation measured by infrared interferometry in the far-field electromagnetic zone. As seen from α-$MoO_3$ 2D crystals, the emission holds well-defined polarization states determined by the material anisotropy. With remarkable agreement, NSSC and NFTE are theoretically described by fluctuation-dissipation theory that permits calculating the HE-LDOS, which is shown to be prime fundamental property ruling near- and far-field radiation emission from 2D hyperbolic crystals. Experiments and theory show that thickness and anisotropy of the medium and temperature consist of tuning parameters that can be used to modulate and control both phenomena. Our work sets path for future research that can address the role of temperature for the near-field superscattering, by taking advantage of experimental setups enabling probing temperature-dependent near-field effects[52,53], and investigate thermal emission from heterostructures formed by different 2D crystals and in twisted architectures following the recent report on mid-IR chiral emission[51] . In summary, by interconnecting the near and the far fields, novel research directions in photonics can rise from the exploitation of NSSC and NFTE using well-established and continuously advancing methods[54,55] to modify the polaritonics of 2D materials in the nanoscale[56].

# Methods

## Scattering-type Scanning Near-field Optical Microscopy (s-SNOM)

The s-SNOM (neaSNOM, attocube systems AG) operates in tapping mode, utilizing a metallized nanoprobe (tip), similar to that of atomic force microscopes (AFM). In our experiments, a quantum cascade laser and a free electron laser are employed as illumination sources. These light sources are focused on tip-sample region, creating intense optical near field created at the tip apex that induces the formation of a local tip-sample effective polarizability. The resultant light scattering from this tip-sample interaction carries the optical near-field response of the sample. This near-field signal is measured by a mercury-cadmium-telluride (MCT) detector in combination with a lock-in based electronic scheme that filters out the background contributions[57]. Such background suppression, basically, consists of demodulating the near-field signal in harmonics of the tapping frequency. The optical near-field signal is recognized by the high harmonic components, typically, beginning from the second harmonic once. The pseud-heterodyne interferometric detection[58] was also applied. The s-SNOM images in the mid-IR were performed at the Imbuia beamline of Sirius, the Brazilian Synchrotron Light Laboratory (LNLS), Brazil, and those in the far-IR, shown in the supplementary materials' section 7, were acquired at the free-electron laser (FEL, supplementary materials' section 6), FELBE at Helmholtz-Zentrum Dresden-Rossendorf, Germany.

## Synchrotron Infrared Nanospectroscopy (SINS)

SINS is performed by a s-SNOM microscope using synchrotron broadband IR radiation to excite the tip-sample interaction[59]. SINS arrangement consists of asymmetric Michelson interferometer where the fixed arm is set by tip-sample and the scanning arm by a gold-coated mirror onto a translation stage. Concisely, the broadband IR near-field light scattered from the tip-sample interaction and the far-field beam from the interferometer scanning arm are collinearly overlapped on a MCT detector. Thus, interferometry is performed by moving the scanning arm. Amplitude and phase SINS spectra are obtained from conventional Fourier transform (FT) of the interferograms. As in s-SNOM imaging, background suppression is achieved by harmonics demodulation through lock-in electronics. In this work, SINS experiments have been done at the Imbuia beamline of Sirius, the Brazilian Synchrotron Light Laboratory (LNLS), Brazil, and at the Advanced Light Source (ALS).

## Infrared Spatial Modulation Spectroscopy (IR-SMS)

Thermal emission spectra from individual a-MoO3 flakes with thickness (d), placed on an Au substrate, were measured using infrared spatial modulation spectroscopy (IR-SMS)[60]. The sample was mounted on a temperature-controlled heating stage and measured under ambient conditions at temperatures up to 450 K. The emitted infrared radiation was collected using a Cassegrain objective with NA = 0.78, directed into a commercial Fourier-transform infrared spectrometer (Bruker), and detected with a mercury cadmium telluride (MCT) detector.

Polarization-resolved spectra were acquired by placing a polarizer (analyser) in the collection path, enabling analysis of the emitted radiation along selected directions with respect to the a-MoO3 crystallographic axes.

To isolate the thermal emission of the a-$MoO_3$ flakes from the background contribution, spatial modulation was implemented by periodically displacing the sample in the object plane using a piezoelectric translation stage. The detector signal was demodulated with a lock-in amplifier at the modulation frequency, thereby suppressing the uniform thermal background and isolating the emission associated with the spatially localized flake. This approach enables far-field detection of thermal emission from micrometre-scale objects.

**COMSOL Simulation**

Numerical simulations were carried out using COMSOL Multiphysics 6.2 with the *Electromagnetic Waves, Frequency Domain* interface in 2D. For the α-$MoO_3$ sample, the modelling approach was based on the developed spheroidal-tip method described in Hu *et. al*[61] . The simulation plane was aligned with the crystallographic axes of α-$MoO_3$, with the [100] direction horizontal and the [010] direction vertical. A metallic spheroidal tip was positioned above the sample surface and illuminated by a *p*-polarized plane wave incident at 60°, generated through periodic ports. Both the excitation frequency and the lateral tip position were scanned: the frequency was swept from 700 to 1100 $cm^{-1}$, and the tip was displaced over a 1 μm distance across the sample. The normal component of the electric field was extracted at the tip apex for each frequency–position pair.

For simulations extending onto the gold substrate, the objective was to isolate the scattered contribution associated with the presence of the α-$MoO_3$ flake. The background field was computed from a *p*-polarized plane wave generated by periodic ports in the absence of the sample, while the total field in the presence of the sample was obtained using the scattered field formulation. The scattered field, defined as the subtraction of these two solutions, was then evaluated along a line starting at the sample edge and extending over 30 μm on the substrate.

**Acknowledgements**

We thank the Brazilian Synchrotron Light Laboratory (LNLS / CNPEM) for providing support to the SINS and s-SNOM nanoimaging in the mid-infrared range that were performed at the IMBUIA beamline through the proposal 20241432. We also thank the Advanced Light Source (ALS) for additional SINS experiments reaching the far-IR spectrum. The Helmholtz-Zentrum Dresden-Rossendorf (HZDR) is thanked for providing beamtime for the nanoimaging experiments at far-infrared frequencies (24203572-ST). FCBM acknowledges support from Fapesp (2022/02901-4) and CNPq (306837/2025-0). A.D is a Research Fellow of the Fonds de la Recherche Scientifique – FNRS. M.L. is a Research Associate of the Fonds de la Recherche Scientifique – FNRS. This research used resources of the "Plateforme Technologique de Calcul Intensif (PTCI)" (http://www.ptci.unamur.be) located at the University of Namur, Belgium, which is supported by the FNRS-FRFC, the Walloon Region, and the University of Namur

(Conventions No. 2.5020.11, GEQ U.G006.15, 1610468, RW/GEQ2016 et U.G011.22). The PTCI is member of the "Consortium des Équipements de Calcul Intensif (CÉCI)" (http://www.ceci-hpc.be). This research used resources of the Advanced Light Source, which is a DOE Office of Science User Facility under contract no. DE-AC02-05CH11231. Parts of this research were carried out at the ELBE Center for High-Power Radiation Sources at the Helmholtz-Zentrum Dresden-Rossendorf e.V., a member of the Helmholtz Association. M.O., F.K., J.W., O.H., L.M.E., and S.C.K. acknowledge the financial support by the Bundesministerium für Bildung und Forschung (BMBF, Federal Ministry of Education and Research, Germany, Project Grant Nos. 05K19ODA, 05K19ODB. and 05K22ODA) and by the Deutsche Forschungsgemeinschaft (DFG, German Research Foundation) under Germany's Excellence Strategy through the Würzburg-Dresden Clusters of Excellence ct.qmat – Complexity and Topology in Quantum MaOer, and ctd.qmat – Complexity, Topology, and Dynamics in Quantum Matter (EXC 2147, project-id 390858490) as well as through the Collaborative Research Center on Chemistry of Synthetic Two-Dimensional Materials-CRC1415 (ID: 417590517).

**Conflict of interest**

Adrian Cernescu is an employee of the company Attocube Systems GmbH that develops and commercializes s-SNOM nanoscopes as the those used in this research.

**Authors Contributions**

FCBM conceived and supervised the project. FHF, FCBM, IDB, YBM, MVP, GN and FB conducted and performed the SINS experiments. FHF, FCBM, YBM, MVP, GN, AC, RAM, WZ, RJ, HW and MLiu conducted and performed ps-het and self-homodyne near-field experiments in the mid-IR using QCL lasers as light source. MB, MLobet and FHF performed analytical and theoretical calculations. AD and ML performed the COMSOL simulations. FHF, FCBM, MO, FK, JW, OH, SW, JMK, LME and SCK performed and conducted near experiments in the mid- to the far-IR using the FEL as light source. AIFTM, GN and YDW performed and conducted IR-SMS experiments. IDB and MVP provided the exfoliated samples. FHF, FCBM, SCK and MLobet drafted the manuscript. All authors contributed to discussions and the review of the manuscript final version.

# Supplementary Materials for:

# Infrared light scattering and emission from epsilon near-zero hyperbolic phonon-polaritons in two-dimensional crystals

Flávio Henrique Feres[1, a], Martin Bearzatto[2], Adrien Debacq[2], Gergely Nemeth[3], Rafael A. Mayer[1], Yuri B. Marçal[1, 4], Marcus V. de Paiva[1,5], Maximilian Obst[6], Felix Kaps[6], Jakob Wetzel[6], Osama Hatem[7], Wenjun Zheng[8], Ran Jing[8], Heng Wang[8], Adrian Cernescu[9], Stephan Winnerl[10], J. Michael Klopf[10], Mengkun Liu[8], Ferenc Borondics[3], Ingrid D. Barcelos[1], Raul O. Freitas[1], Ana I. F. Tresguerres-Mata[12], Yannick De Wilde[12], Lukas M. Eng[6, 13], Susanne C. Kehr[6], Michaël Lobet[2, b] & Francisco C. B. Maia[1, c]

14 Brazilian Synchrotron Light Laboratory (LNLS), Brazilian Center for Research in Energy and Materials (CNPEM), 13083-970, Campinas, São Paulo, Brazil
15 University of Namur, Department of Physics and Namur Institute of Structured Material, 61 rue de Bruxelles, 5000 Namur, Belgium
16 SMIS Beamline, SOLEIL Synchrotron, L'Orme des Merisiers, RD128, 91190 Saint Aubin, France
17 Physics Department, Gleb Wataghin Physics Institute, University of Campinas (Unicamp), Campinas, SP, Brazil
18 Department of Physics, Federal University of São Carlos 13565-905, Sao Carlos, SP, Brazil
19 Institute of Applied Physics, Technische Universität Dresden, Dresden, Germany
20 Departament of Engineering Physics, Faculty of Engineering, Tanta University, Tanta 31511, Egypt
21 Department of Physics and Astronomy, Stony Brook University; Stony Brook, New York 11794, USA
22 attocube systems GmbH, 85540, Haar-Munich, Germany.
23 Institute of Ion Beam Physics and Materials Research, Helmholtz-Zentrum Dresden-Rossendorf (HZDR), Dresden, Germany
24 Institute of Radiation Physics, Helmholtz-Zentrum Dresden-Rossendorf (HZDR), Dresden, Germany
25 Institut Langevin, ESPCI Paris, PSL University, CNRS, Paris, France
26 Würzburg – Dresden Cluster of Excellence (EXC2147) ctd. qmat – Complexity, Topology, and Dynamics in Quantum Matter, Dresden, Germany

**Corresponding authors:**

d. flavio.feres@lnls.br;
e. michael.lobet@unamur.be;
f. francisco.maia@lnls.br

# Index

# 1 Near-field superscattering from hexagonal boron nitride two-dimensional crystals

The SINS linescan, shown in **Fig. SM1a**, reveals the near-field superscattering effect, around 820 cm$^{-1}$, from a 150 nm-thick hBN crystal lying on Au. In this measurement, the nanoprobe scanned a total path of 35 µm away from the crystal on the Au surface: the crystal extended from x = 0 to 3 µm and the remaining of the scan was performed on the Au substrate. As discussed for $\alpha$-$MoO_3$ in the main text, the superscattering rises at the material's ENZ range of the $\varepsilon_{001}$ component since $\varepsilon_{001}$~ 0 at $\omega_{001}^{LO}$= 820 cm$^{-1}$. This range is zoomed in **Fig. SM1b** that permits observing the oscillatory characteristic of the NSSC radiated from the crystal. Such wavy behaviour is seen from the spatial profiles presented in **Fig. SM1c**, which correspond to the different frequencies indicated by the white horizontal dashed lines in **Fig. SM1b**. Moreover, in **Fig. SM1d** upper panel we present the nanoimage of the superscattering from a 73.5 nm-thick hBN, residing on Au substrate, using the free-electron laser (FEL), FELBE at Helmholtz-Zentrum Dresden-Rossendorf, Germany, tuned at 820 cm$^{-1}$ as illumination source. As seen in corresponding spatial profile (**Fig. SM1d lower panel**), the effect extends over 40 µm far from the crystal.

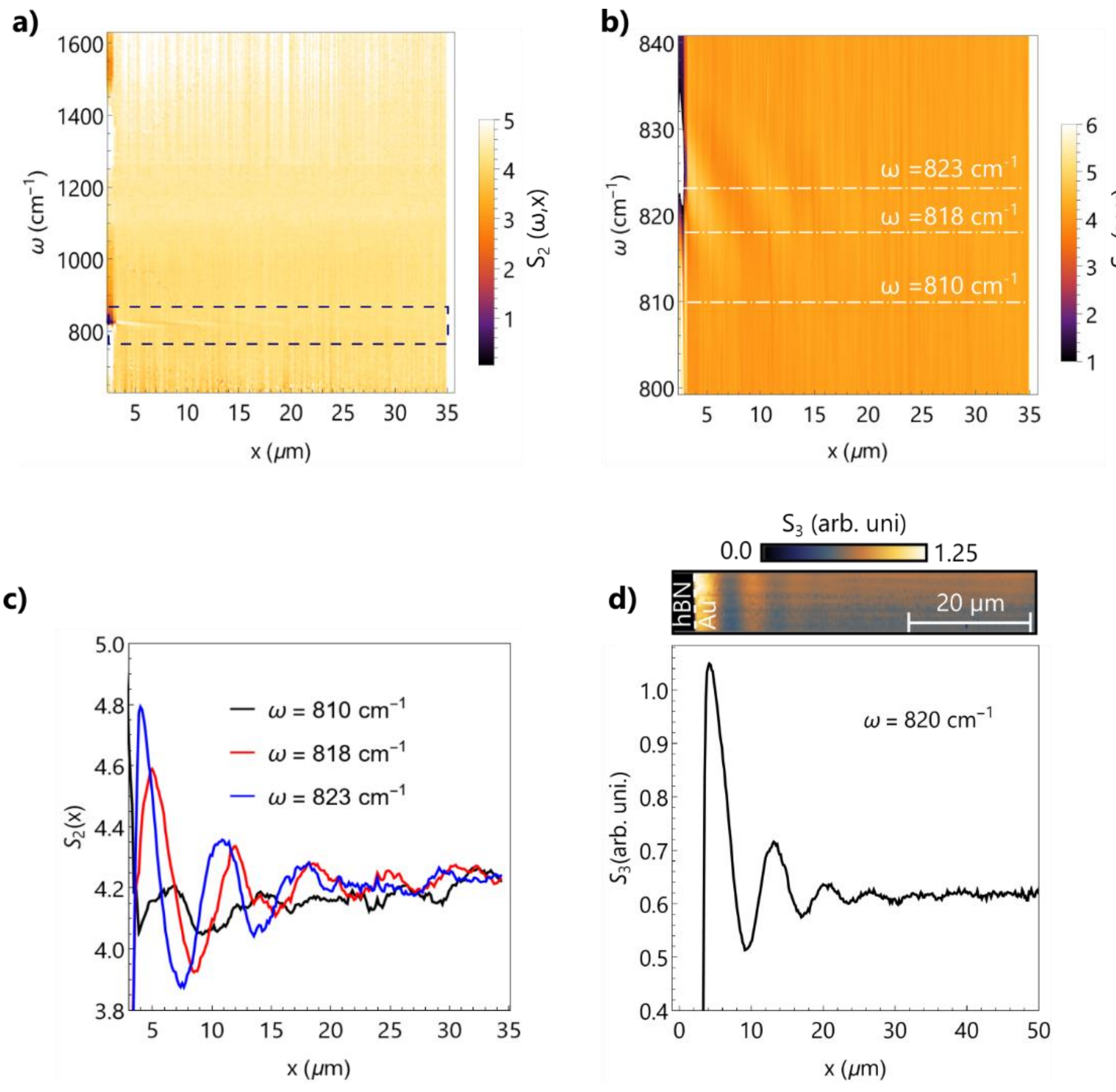

Fig SM1. **a)** Second harmonic amplitude, $S_2(\omega, x)$, SINS linescan over 35 µm: in 0 ≤ x ≤ 3 µm the nanoprobe scans a 150 nm-thick hBN and, in $3 < x \leq 35\ \mu m$, the Au surface. **b)** Zoom of the measurement presented in **(a)** for the ENZ range of hBN. The white horizontal dashed lines indicate NSSC spatial profiles, extracted for different $\omega$'s that are shown in **(c)**. Upper panel of **(d)** displays the FEL nanoimage at $\omega$ = 820 cm$^{-1}$ of the NSSC effect from a 73.5 nm-thick hBN with the corresponding spatial profiles presented in the lower panel of **(d)**.

## 2 Mie Scattering Theory for a cylindric and anisotropic material

To describe the scattering response of the α-$MoO_3$ crystals, we employed a generalized two-dimensional Mie formalism for a cylindrical scatterer embedded in an anisotropic medium[2]. The incident field ($\boldsymbol{E_0}$) is described in cylindrical harmonics using Bessel functions $\boldsymbol{J_m}(k\rho)$, whereas the scattered field is represented in terms of outgoing Hankel functions $\boldsymbol{H_m^{(1)}}(k\rho)$. The total field was then calculated as the coherent superposition of the incident and scattered waves

$$E_z(\rho,\varphi) = E_{z0} + E_z^{sct} = E_{z0} + E_{z0}\sum_m i^m a_m H_m^{(1)}(k\rho)e^{im\varphi} \quad \text{Eq. SM3.}$$

The anisotropic dielectric response of α-$MoO_3$ was incorporated through the tensorial permittivity and the corresponding hyperbolic dispersion relation[3] . The scattering coefficients $\boldsymbol{a_m}$ were obtained by imposing the continuity of the tangential electromagnetic fields at the crystal boundary

$$a_m = \frac{k_2 J_m'(Z_a) J_m(X_a) - k_1 J_m(Z_a) J_m'(X_a)}{k_1 J_m(Z_a) H_m^{(1)\prime}(X_a) - k_2 J_m'(Z_a) H_m^{(1)}(X_a)} \quad \text{Eq. SM4}$$

Where $Z_a = k_2 a$ , $X_a = k_1 a$, $k_2 = \sqrt{\varepsilon_{xx}\left(k_0^2 - \frac{q^2}{\varepsilon_{zz}}\right)}$ and $k_1 = k_0 \cos\theta$.

Using this approach, we generated the scattering field structures considering an infinite and cylindrical-like α-$MoO_3$ crystals of different radii a (**Fig. SM6a-c**).

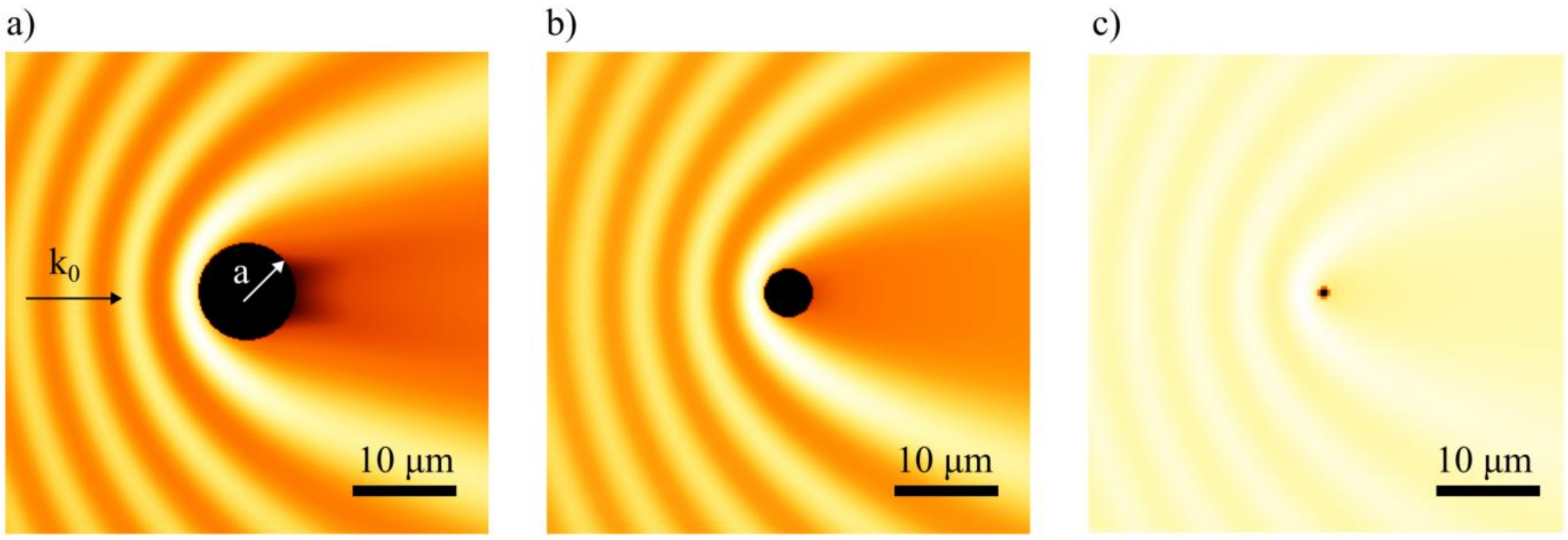

Figure SM6. Maps of the electric field amplitude, | $E_z$ |, around a dielectric cylinder under plane-wave illumination. (a) $a = 5\ \mu m$, (b) $a = 2.5\ \mu m$ and (c) $a = 1\ \mu m$. Scale bars: 10 $\mu$m

Within the framework of Mie scattering theory [4], the scattered field can be expressed as a superposition of cylindrical harmonics, where the scattering amplitudes are determined by the modal reflection coefficient ($R_m$) of the supported polaritonic modes ($a_m = \frac{R_m - 1}{2}$). The total scattering cross section ($\sigma_{sc}$) is obtained by summing the contributions of all scattering channels and is therefore governed by the magnitude of the corresponding reflection coefficients.

$$\sigma_{sc} = \frac{4}{k} \sum_{m=-\infty}^{+\infty} |a_m|^2 = \frac{1}{k} \sum_{m=-\infty}^{+\infty} |R_m - 1|^2 \qquad \text{Eq. SM5}$$

Assuming that $R_m$ in eq. SM5 can be roughly represented by the $p$-polarized reflection coefficient obtained from solving Maxwel equations for an infinite α-$MoO_3$ slab, in **Fig. SM7** we show that the calculated $\sigma_{sc}$ spectrum for the NSSC, considering the integration over different momentum orders $l$ within the range from $q = 0$ to $8.0 \times 10^5$ cm$^{-1}$, surpasses the scattering at the single channel limit.

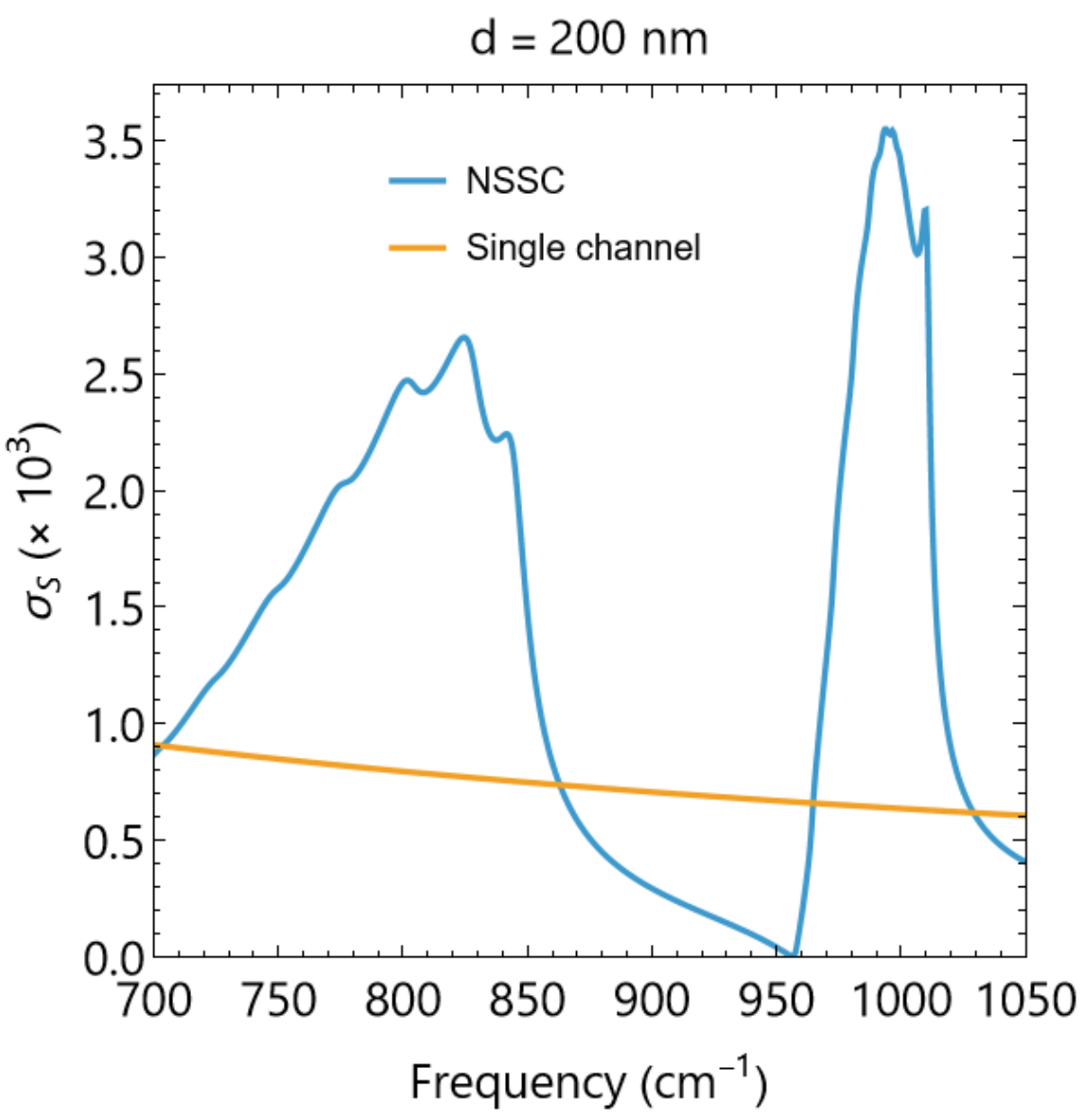

Fig. SM7 – Calculation of the scattering coefficients of the NSSC (Eq. SM5) and the single channel limit.

In hyperbolic α-$MoO_3$, each HPhP branch acts as an independent polaritonic scattering channel. Near the ENZ-HPhP frequency, multiple polaritonic channels exhibit enhanced reflection amplitudes simultaneously, resulting in a cumulative increase of the $\sigma_{sc}$. The contribution of these channels gives rise to a superscattering regime characterized by an exceptionally strong scattering response. Since the same polaritonic resonances also govern the enhancement of the photonic local density of states (LDOS), both superscattering and LDOS enhancement originate from the same underlying hyperbolic polaritonic modes.

## 3 Frequency ($\omega$) -momentum ($q$) dispersion of hyperbolic phonon-polaritons

In Fig. SM2, it is shown the calculated $\omega - q$ dispersion of HPhP modes, in the upper hyperbolic band, of a 200 nm-thick $\alpha$-$MoO_3$. As commented in the discussion of the superscattering from the measurements shown in **Fig. 1e-j** of the main text, it is displayed in Fig. SM2 that the extracted HPhP momenta from the 995 and 1005 $cm^{-1}$ profiles (black points) match the theoretical prediction from $Im(r_p)$.

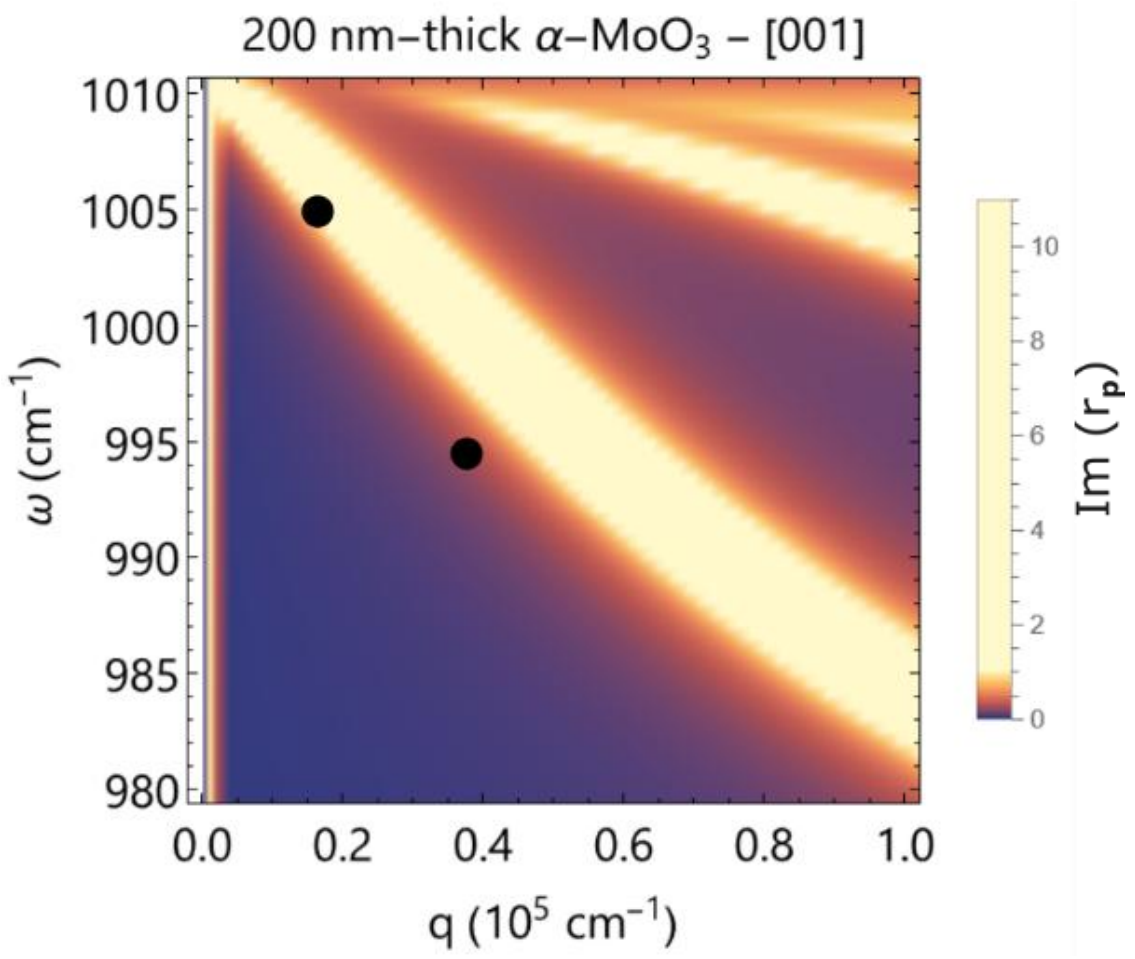

Fig SM2. Hyperbolic Phonon Polariton Dispersion of 200 nm thick crystal lying on Si substrate. The data points were extracted from the profiles in **Fig. 1f** and **1h** of the main text.

## 4 Near-field superscattering from molybdenum trioxide in mid- and far-infrared

We show in **Fig. SM3a** the $S_2(\omega, x)$ SINS linescan, extending over 35 μm from the edge of a 808 nm-thick α-$MoO_3$ crystal to the Au substrate. The measurement covers the 350 – 1100 $cm^{-1}$ broad spectral range, covering the far- to mid-infrared ranges. In **Fig. SM3b** we present spatial profiles of NSSCs associated with the ENZ ranges of different hyperbolic bands (HB) of the material. In this case, the 500 $cm^{-1}$ and the 1000 $cm^{-1}$ profiles are related to the ENZ ranges of the type I HBs at $\omega$ = 444 – 508 $cm^{-1}$ and $\omega$ = 956.7 – 1006.9 $cm^{-1}$ and, the 822 $cm^{-1}$ profile can be ascribed to the ENZ of the type II HB at $\omega$ = 544.6 – 850 $cm^{-1}$. In **Fig. SM3c**, the α-$MoO_3$ hyperbolic bands can be spectrally localized in the frequency intervals where the real parts of its electrical permittivity components assume negative values[1]. Such bands rise due to phonon resonances of the α-$MoO_3$ electrical permittivity components[1] (**Fig. SM3c**) which are ascribed to: (i) $\varepsilon_{cc}$ ([010] axis) phonon resonances at $\omega$ = 444 - 508 $cm^{-1}$ and $\omega$ = 956.7 - 1006.9 $cm^{-}$; (ii) $\varepsilon_{bb}$([001] axis) phonon resonances at $\omega$ = 544.6 - 850 $cm^{-1}$; and (iii) $\varepsilon_{aa}$ ([100] axis) phonon resonances at $\omega$ = 506.7 - 534.3 $cm^{-1}$, $\omega$ = 821.4 - 963 $cm^{-1}$ and $\omega$ = 998.7 - 999.2 $cm^{-1}$. Furthermore, in **Fig. SM3d** we present the superscattering of 320 nm-thick crystal measured by s-SNOM nanoimaging using FEL illumination at 500 $cm^{-1}$. In analogy to the image at 1002 $cm^{-1}$ described in **Fig. 2e**, we observe that the 500 $cm^{-1}$ signal is minimal on the crystal and the scattering field structure is asymmetric with respect to the crystal position. As commented, this asymmetrically structured optical field forms due the complex interference between the radiating polariton fields from the crystal and far-field components that reflect at the crystal. The Mie scattering approach, discussed in the main text and in section **2**of this document, can reliably reproduce the measured NSSC.

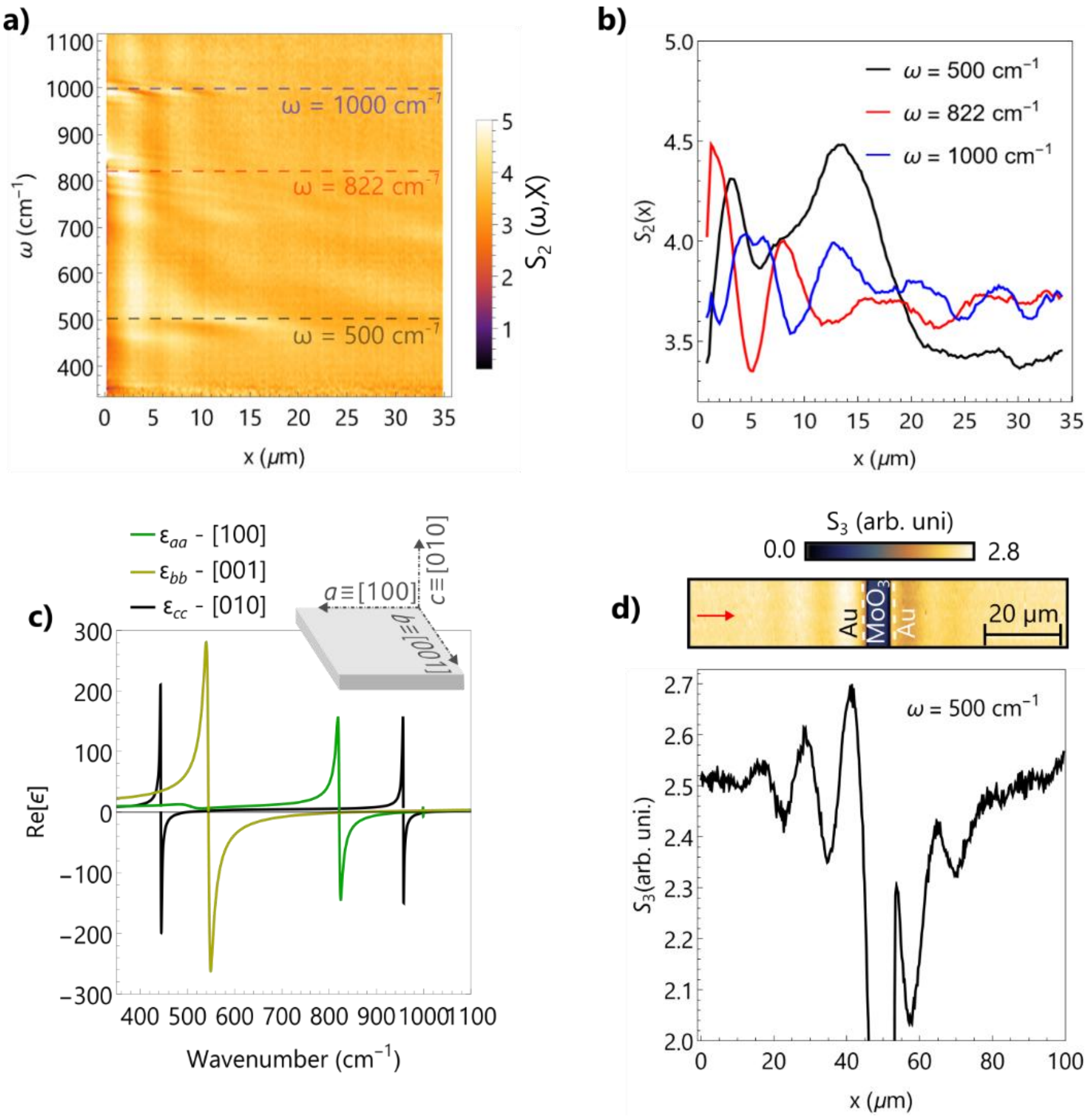

Figure SM3. a) $S_2(\omega, x)$ SINS linescan, extending over 35 μm from the edge of a 808 nm-thick α-$MoO_3$ crystal deposited on the Au substrate. b) NSSC profiles extracted from the measurement in (a) at the frequencies denoted by dashed lines. c) Real parts of the α-$MoO_3$ electrical permittivity components[1] . d) Upper panel, s-SNOM nanoimage of the NSSC from a 320 nm-thick α-$MoO_3$ illuminated at 500 $cm^{-1}$ and, bottom panel, the corresponding superscattering profile.

## 5 Near-field superscattering central frequency and linewidth

**In Fig. SM5a**, we plot the NSSC spectra for α-$MoO_3$ crystals of different thicknesses $d$. The data (symbols) consist of the measured spectra at $x \sim 600$ nm in the linescan measurements which are shown in **Fig. 4a-c** of the main text. The spectra indicate a broadening of the superscattering as function of $d$. In good agreement, the calculated HE-LDOS, the same thicknesses (**Fig. SM5b**), present the same trend. To quantitatively examine the NSSC, we use the complex Lorentzian-model (eq. SM1) to describe the measured amplitude of the NSSC spectrum. In this expression, the index $m$ corresponds to the number of peaks and, $A_i$, $\omega_{c,m}$ and $\gamma_m$ are the peak amplitude, the central frequency and the full width at the half maximum of each peak, respectively. $b$ and $\phi$ are, respectively, the amplitude and phase of a complex non-resonant background. The fit-extracted $\omega_{c,m}$'s and $\gamma_m$'s are listed in Table 1 and compared with the corresponding predictions, from HE-LDOS theory, in Fig. SM4c and 4d, respectively.

$$s(\omega) = \left| \sum_m \frac{A_m.\omega.\gamma_m}{\left(\left(\omega^2 - \omega_{c,m}{}^2\right) - i.\omega.\gamma_m\right)} + be^{i\phi} \right| \qquad \text{Eq. SM1}$$

*Table 1. Fitting parameters from the eq. SM1 model.*

| | $d = 140\ nm$ | | $d = 270\ nm$ | | $d = 808\ nm$ | |
|---|---|---|---|---|---|---|
| $\gamma_1$ | 3.07 | ± 0.19 | 6.72 | ± 0.37 | 7.5 | ± 1.3 |
| $\omega_{c,1}$ | 1001.9 | ± 0.1 | 999.7 | ± 0.2 | 991.6 | ± 0.5 |
| $\gamma_2$ | - | | - | | 5.7 | ± 2.3 |
| $\omega_{c,2}$ | - | | - | | 1005.8 | ± 0.7 |

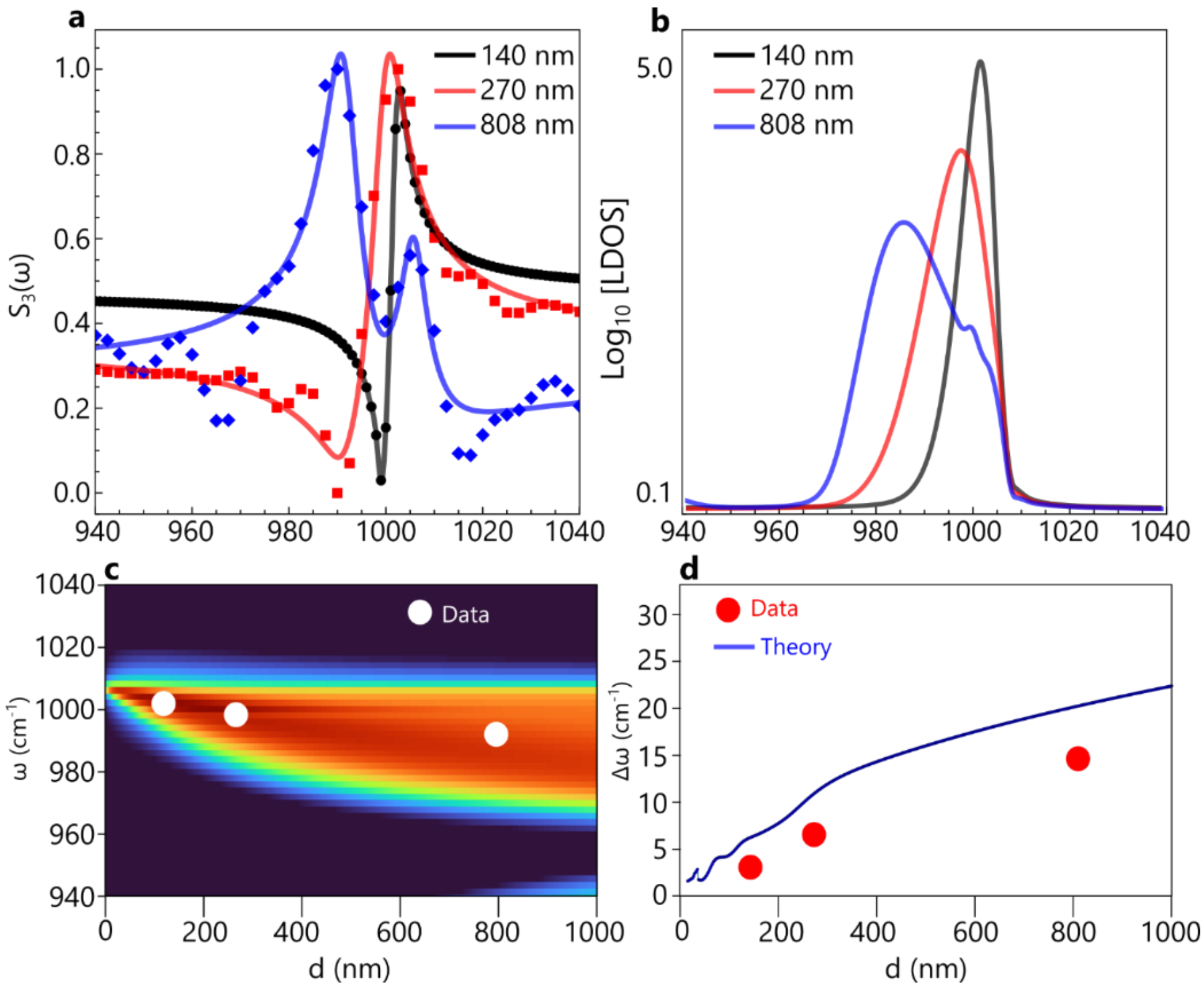

Figure SM5. (a) NSSC spectra extracted close to the interface for α-$MoO_3$ crystals of thicknesses $d$ = 140 nm (black), 270 nm (red) and 808 nm (blue). The maxima of the data are normalized to 1. The symbols represent the data points and the solid line the Lorentzian fitting. (b) HE-LDOS calculated at 50 nm from the interface for the same thicknesses. (c) HE-LDOS as a function of thickness $d$. The white symbols are central frequency data extracted from the fitting in a). (d) Spectral linewidth extracted from the FWHM of the HE-LDOS (blue) and from fitting the experimental data (red). The data point for d = 808 nm is the sum of the corresponding $\gamma_1$ and $\gamma_2$.

## 6 In-Plane Angle Dependence Measurements.

Narrowband near-field images acquired on an α-$MoO_3$ crystal deposited on Au, revealing the superscattering response. The measurements were performed at the HZDR Free-Electron Laser facility (Dresden) using a FEL-coupled s-SNOM system. The scale bars correspond to 5 µm. The bottom panels show the scattering response for different azimuthal orientations of the crystal with respect to the incident beam.

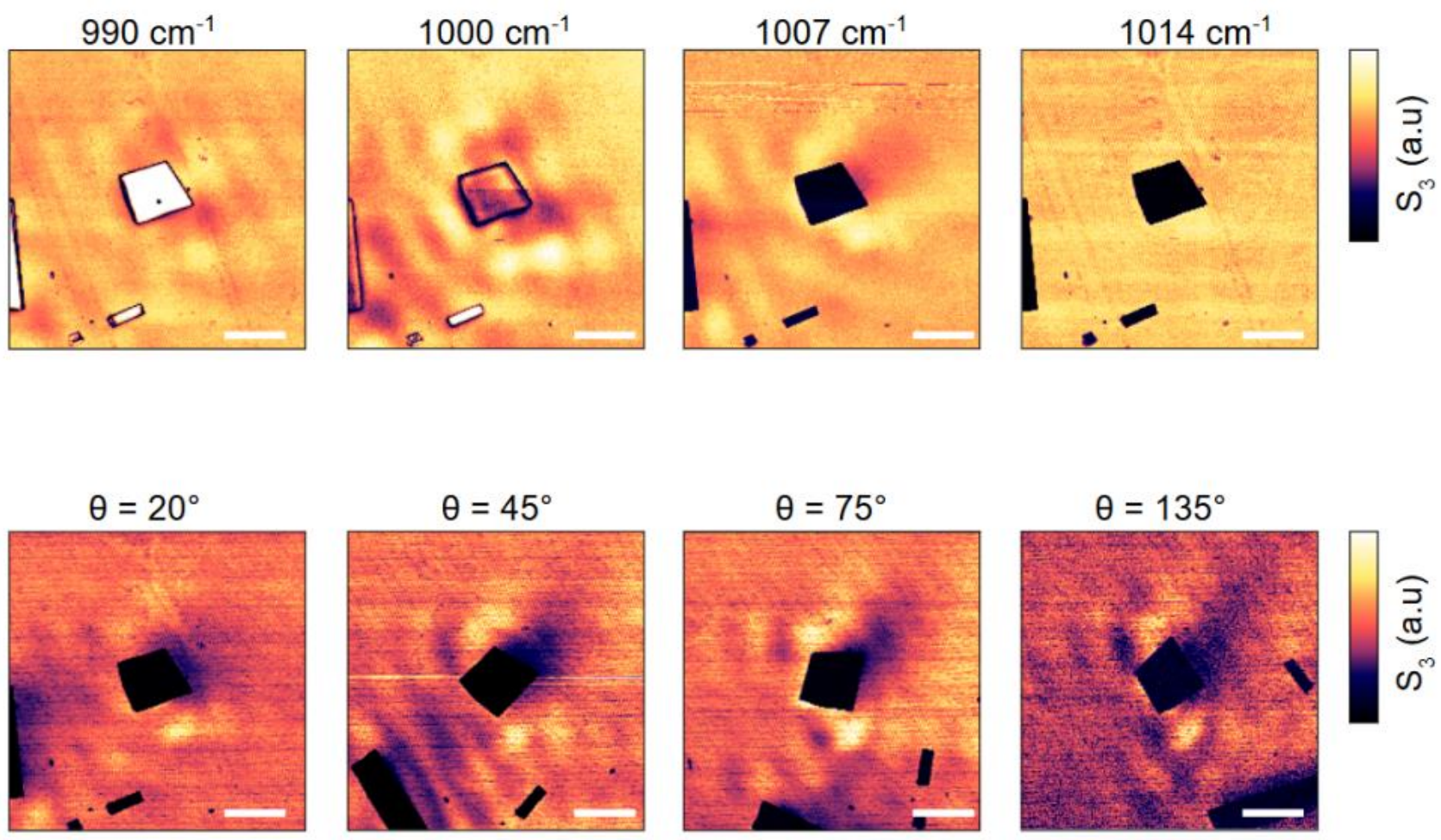

Figure SM4. s-SNOM images of an α-$MoO_3$ crystal on a gold substrate, acquired FEL excitation. The top panels show the frequency-dependent scattering response recorded across the ENZ-HPhP spectral region at 990, 1000, 1007, and 1014 $cm^{-1}$. The bottom panels display measurements performed at different azimuthal orientations (θ = 20°, 45°, 75°, and 135°) of the crystal relative to the incident FEL beam.

## 7 Near-field thermal emission dependence on crystal thickness

The dependence on the thickness *d* of the near-field thermal emission is experimentally demonstrated for two α-$MoO_3$ with d = 150 nm (**Fig. SM8a**) and d = 500 nm (**Fig. SM8b**). The NFTE spectral data of both crystals present two bands around 800 and 1000 $cm^{-1}$ that increase with the temperature as predicted by theory. The overall intensity of the thicker crystal is about 5-to-6-fold higher, or even about 10-fold for the 800 $cm^{-1}$ peak, than that of the thinner one. However, in the 500 nm-thick crystal spectrum the 800 $cm^{-1}$ band shows a higher amplitude than the 1000 $cm^{-1}$ one, whilst the opposite relation is observed in the spectrum for the 150 nm-thick crystal.

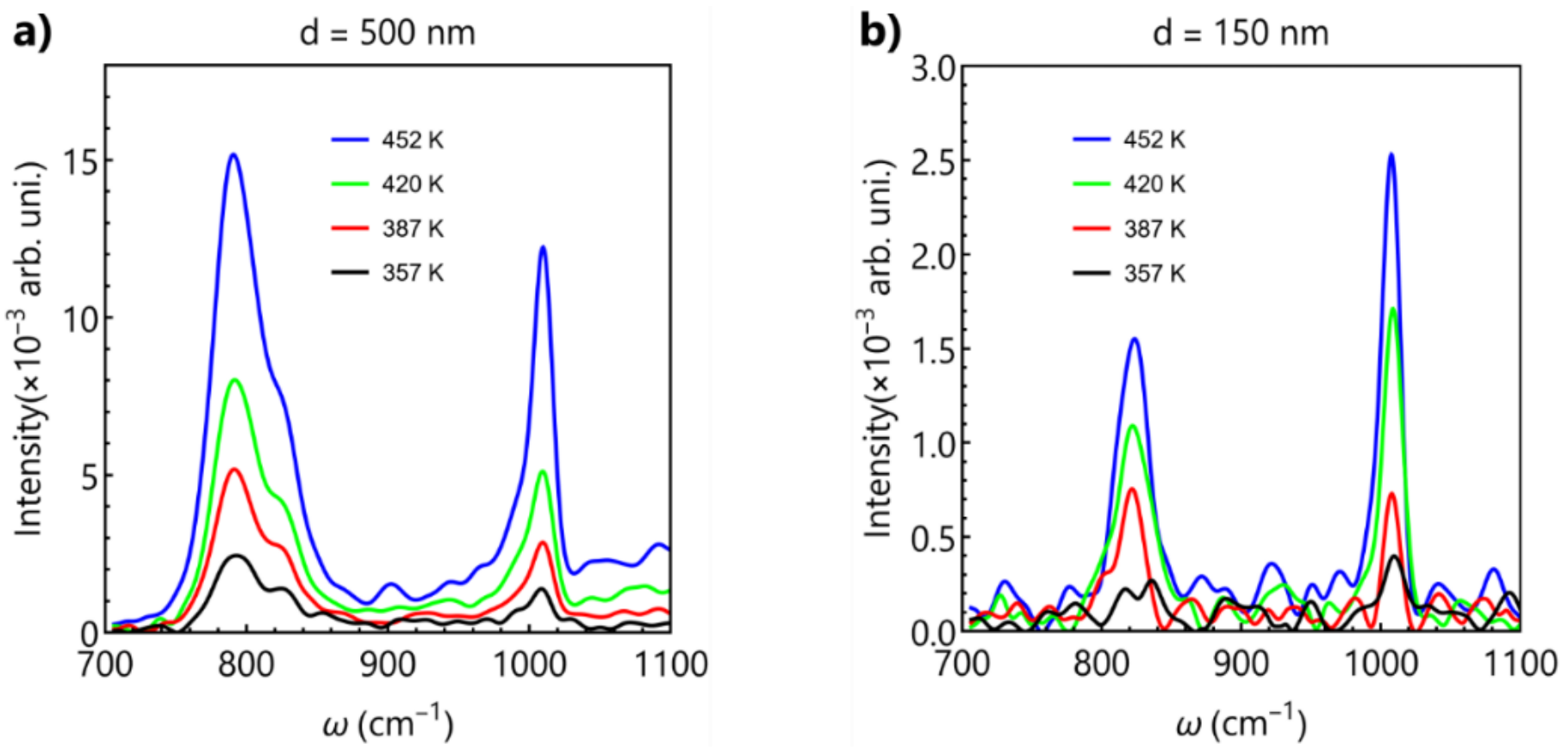

Figure SM8. Near-field thermal emission (NFTE) spectra measured on α-$MoO_3$ crystals with different thicknesses: **(a)** *d* = 500 nm and **(b)** *d* = 150 nm. Spectra were acquired at different temperatures ranging from 357 K to 452 K.

# 8 Near-field thermal emission linewidth

The NFTE from a 500 nm-thick α-$MoO_3$ crystal, measured with a spectral resolution of 2 $cm^{-1}$, is displayed in **Fig. SM9**. To determine the linewidth of the band around 1000 $cm^{-1}$, we use a gaussian model $I(\omega)$ (eq. SM2) that computes the summation over components $m$. In analogy to the modelling the section 5 of this supplementary materials document, the eq. SM2's parameters are: amplitude $A_m$, spectral broadening $\Upsilon_m$, central frequency $\Omega_m$ and non-resonant background $C$. The model-fit of such band is formed by two components centred at $\Omega_2 = 1000.1$ and $\Omega_1 = 1005.6$ $cm^{-1}$. The 1005.6 peak has higher amplitude and spectral broadening $\Omega_2 = 3$ $cm^{-1}$ that is close to the spectral resolution used in the IR-SMS measurements. All fit-extracted parameters are presented in Table 2. To calculate the quality facto $Q$, we use the relation $Q = {}^{\Omega}/_{2\sqrt{ln2}.\Upsilon}$ since the term in the denominator represents the full width at half maximum (FWHM). Using the fit-extracted values for the 1005.6 cm-1 peak, we obtain $Q = 201$ as discussed in the main text.

$$I(\omega) = \sum_m \frac{A_m}{\Upsilon_m} . e^{-\left(\frac{\omega - \Omega_m}{\Upsilon_m}\right)^2} + C \qquad \text{Eq. SM2}$$

*Table 2. Fitting parameters from the eq. SM2 model.*

| | | |
|---|---|---|
| $A_1$ | 43 | ± 2 |
| $\Upsilon_1$ | 3.0 | ± 0.1 |
| $\Omega_1$ | 1005.6 | ± 0.06 |
| $A_2$ | 49 | ± 3 |
| $\Upsilon_2$ | 8.9 | ± 0.4 |
| $\Omega_2$ | 1001.1 | ± 0.4 |

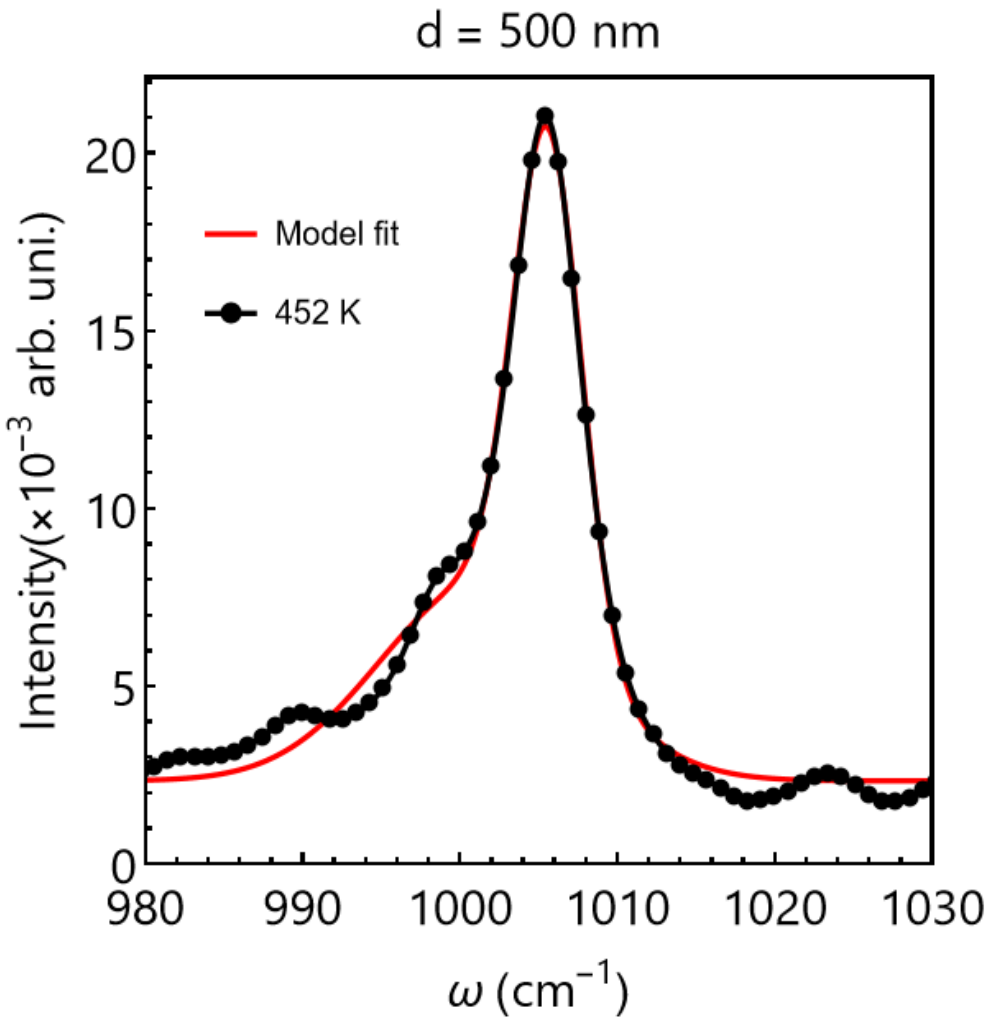

Figure SM9. Model fit curve (red), using eq. SM2, and experimental data of the 1000 $cm^{-1}$ NFTE band measured with 2 $cm^{-1}$ spectral resolution. These data were extracted from Fig. 4c of the main text.

## 9 Impact of film thickness on emitted energy density

The analytical model describing NSSC (Eq 1. of main text) relies exclusively on Fresnel reflection coefficients to represent the LDOS above the interface of an $\alpha$-$MoO_3$ slab infinite along the plane. For both polarisations, at the interface between homogenous isotropic media $i$ and $j$, those coefficients are given by Álvarez-Pérez, G. *et al* [5]:

$$r_{ij}^{s} = \frac{k_{zi}-k_{zj}}{k_{zi}+k_{zj}}, \quad \text{Eq. SM3a,}$$

$$r_{ij}^{p} = \frac{\varepsilon_j k_{zi}-\varepsilon_i k_{zj}}{\varepsilon_j k_{zi}+\varepsilon_i k_{zj}}, \quad \text{Eq. SM3b,}$$

$$t_{ij}^{s} = \frac{2k_{zi}}{k_{zi}+k_{zj}}, \quad \text{Eq. SM3c,}$$

$$t_{ij}^{p} = \frac{2\sqrt{\varepsilon_i\varepsilon_j}k_{zi}}{\varepsilon_j k_{zi}+\varepsilon_i k_{zj}}, \quad \text{Eq. SM3d.}$$

To consider multiple reflections occurring into the slab of finite thickness $d$, the global reflection coefficient of the slab system can be represented using an infinite series for $r_s$ and $r_p$ on a superstrate-slab-substrate system (indexed 1-2-3)[5]:

$$r^{\nu} = r_{12}^{\nu} + \frac{t_{12}^{\nu}t_{21}^{\nu}r_{23}^{\nu}e^{2ik_{z2}d}}{1-r_{21}^{\nu}r_{23}^{\nu}e^{2ik_{z2}d}}, \quad \text{Eq. SM4,}$$

with the polarisation $\nu = s$ or $p$. The $k_{zi}$ components are the wavevector components perpendicular to the interfaces for an incident wavevector $k_0$ and is defined as $k_{zi} = \sqrt{\varepsilon_i\, k_0^2 - k_{\parallel}^2}$ for isotropic media $i$. However, in the case of uniaxial anisotropic media, such as $h$-BN, the definition needs to be adjusted to take into account the in-plane $\varepsilon_{\parallel}$ and out-of-plane $\varepsilon_{\perp}$ permittivities such as $k_z = \sqrt{\varepsilon_{\parallel}\, k_0^2 - (\varepsilon_{\parallel}/\varepsilon_{\perp})k_{\parallel}^2}$ for p-polarisation and $k_z = \sqrt{\varepsilon_{\parallel}\, k_0^2 - k_{\parallel}^2}$ for s-polarisation[5,6] . In the case of a biaxial crystal, that is the $\alpha$-$MoO_3$, the in-plane component of the permittivity needs to be defined as a combination of the in-plane component, $\varepsilon_{\parallel} = \varepsilon_{xx}\cos^2(\alpha) + \varepsilon_{yy}\sin^2(\alpha)$[5,6], with $\alpha$ being the angle between the *x*-axis of the crystal and the in-plane component of the incident wave-vector and the out-of-plane permittivity $\varepsilon_{\perp} = \varepsilon_{zz}$. The permittivity tensor components for $\alpha$-$MoO_3$ are model with a set of Lorentz oscillators[1]. The permittivity used in the definition of the Fresnel coefficients is the in-plane component $\varepsilon_{\parallel}$ exepct in the definition of the wavevector out-of-plane component. This description of the global reflexion coefficient has been validated by computing the dispersion relation of the system and comparing it to the analytical expression[7].

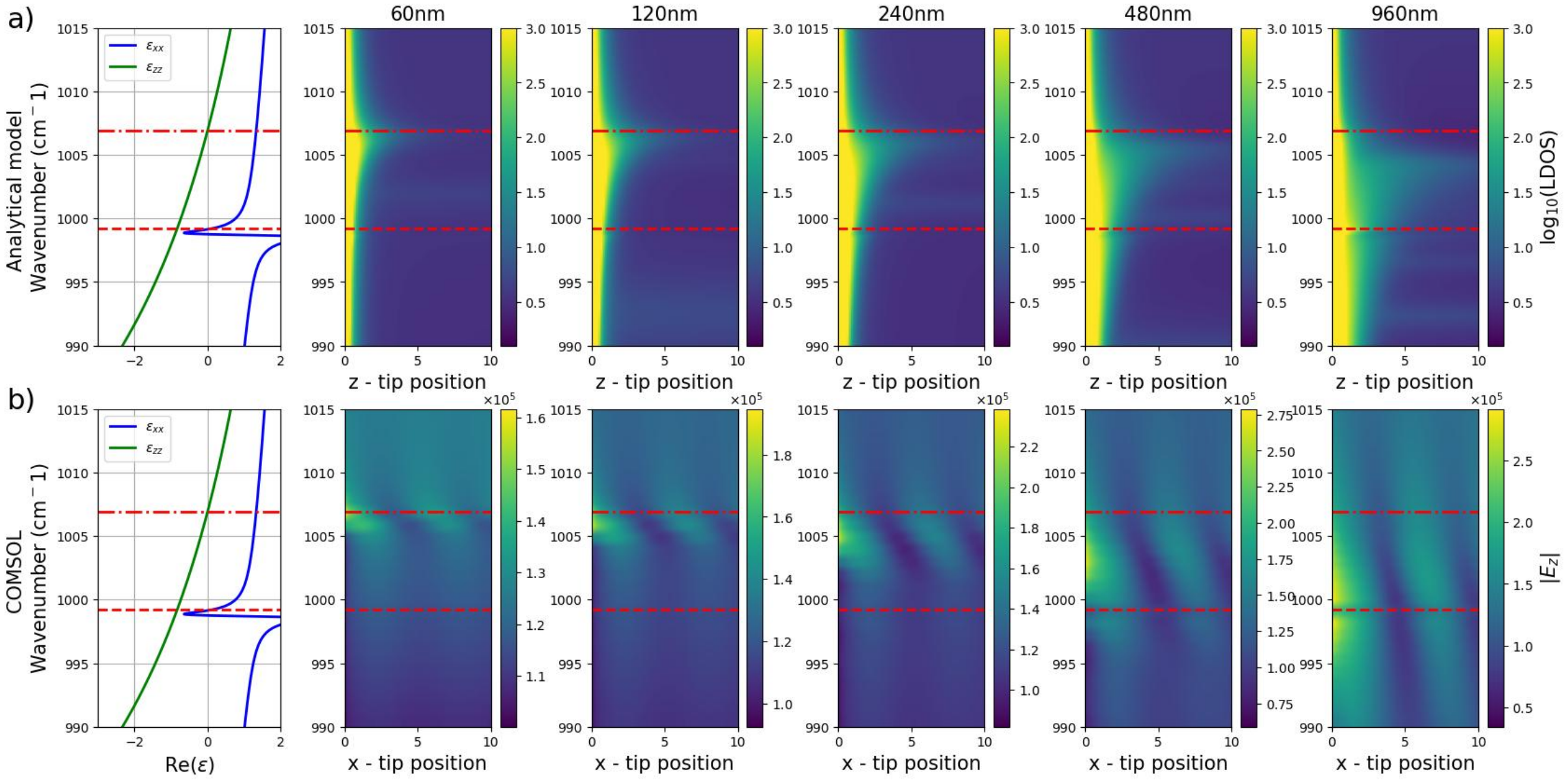

Figure SM10. a) Evolution of LDOS exaltation along z-axis and b) simulated electric field amplitude with increasing thickness of $\alpha$-$MoO_3$ slab on gold substrate. With increasing thickness, the maximum of LDOS (a) is clearly shifted from the ENZ frequency, 1006 $cm^{-1}$ (dash-dotted red line) to lower frequency. The same behavior is presented in the amplitude of the simulated electric field (b) computed using the finite element method implemented in the software COMSOL Multiphysics. Moreover, a drop of intensity is present around 999$cm^{-1}$ (red dashed line), due to the permittivity tensor component $\varepsilon_{xx}$ crossing zero at that frequency, resulting on a narrow wavenumber window where both in- and out-of-plane components of slab permittivity are negative resulting in a vacancy of mode in this region. LDOS of figure (a) was computed using the parameter $\alpha = 0$.

As can be seen in first line of Figure SM10, the model highlights the shift of the LDOS intensity above the slab from the ENZ frequency ($\varepsilon = 0$) to the SPhP frequency ($\varepsilon = -1$) as thickness increases. An explanation for this phenomenon is the fact that for a slab between two media, if the slab thickness is small enough, the surface waves on both interfaces of the slab can interact and the dispersion relation for such surface waves can then split into two modes, introducing an odd and an even mode respectively at higher and lower frequencies than the single mode at a simple interface[8]. These two modes spectrally separate more with decreasing slab thickness, with the odd mode plateauing at the ENZ frequency, i.e. the frequency where Re(ε) = 0, when the thickness of the slab satisfies $d \leq \lambda_p/50$ with $\lambda_p$ the wavelength of the ENZ frequency [8–10]. Moreover, the LDOS spectrum broadens and covers the Reststrahlen band, as mentioned in the main text, resulting from the increase of the number of modes available with the thickness of the slab.

Those results from this analytical model are replicated using COMSOL Multiphysics software (second line of Figure SM8) to model the electric field surrounding an $\alpha$-$MoO_3$ flake, thus showing the robustness of the model and the link between the LDOS and the exaltation of the electric field. The drop in intensity present at 999 $cm^{-1}$ originates from a weak oscillation in the $\varepsilon_{xx}$ permittivity component. The real part of this permittivity component become negative for a narrow frequency range, resulting in a vacancy of mode in this spectral window. Similar results are observed in the LDOS calculated by the analytical model in figure SM8a when using the parameter $\alpha = 0$.